\documentclass[letterpaper,twocolumn,10pt]{article}
\usepackage{usenix}

\usepackage{tikz}
\usepackage{amsmath}

\usepackage{filecontents}

\usepackage{booktabs}
\usepackage{tikz}
\usepackage{comment}
\usepackage{amsmath}
\usepackage{amsfonts}
\usepackage{caption}
\usepackage{subcaption}
\usepackage{graphicx}
\usepackage{float}
\usepackage{tabularx}
\usepackage{algorithm}
\usepackage{algorithmicx}
\usepackage[noend]{algpseudocode}

\algrenewcommand\textproc{}
\algnewcommand{\LineComment}[1]{\State \(//\) #1}
\usepackage[normalem]{ulem}
\usepackage{bbding}
\usepackage{todonotes}
\usepackage{filecontents}
\usepackage{minted}
\usepackage{enumitem}
\usepackage[bottom]{footmisc}
\usepackage{xurl} 
\usepackage{hyperref}
\usepackage{cleveref}
\usepackage{pifont}
\usepackage{multirow}
\usepackage[dvipsnames,svgnames]{xcolor}
\usepackage{outlines}
\newcommand{\sysname}{{MuxTune}}

\newtheorem{lemma}{Lemma}
\newtheorem{theorem}{Theorem}

\makeatletter
\renewcommand{\paragraph}{%
  \@startsection{paragraph}{4}%
  {\z@}{1ex \@plus 1ex \@minus 1ex}{-1em}%
  {\normalfont\normalsize\bfseries}%
}

\begin{document}

\title{\sysname{}: Efficient Multi-Task LLM Fine-Tuning in Multi-Tenant Datacenters \\ via Spatial-Temporal Backbone Multiplexing}
\date{}

\author{
{\rm Chunyu Xue$^{\dagger*}$, Yi Pan$^{\dagger*}$, Weihao Cui$^{\dagger\S}$, Quan Chen$^{\dagger}$, Shulai Zhang$^{\dagger}$, Bingsheng He$^{\S}$, Minyi Guo$^{\dagger}$} \\
$^{\dagger}$Shanghai Jiao Tong University \quad $^{\S}$National University of Singapore \\ 
} 

\maketitle

\begin{abstract}

Parameter-Efficient Fine-Tuning (PEFT) is widely applied as the backend of fine-tuning APIs for large language model (LLM) customization in datacenters.
Service providers deploy separate instances for individual PEFT tasks, giving rise to prominent resource inefficiencies, including (1) GPU underutilization from small-scale, PEFT-native operators and (2) device stalls from communication delays and data dependencies in parallelized execution. 
To address these issues, this paper presents \sysname{}, a fine-tuning system that enables resource-efficient concurrent execution of multiple PEFT tasks. 
The key idea is to multiplex the backbone across independent tasks in a spatial-temporal manner for improved utilization and reduced stalls.
Building on flexible, modularized backbone sharing via unified PEFT representations, \sysname{} proposes hierarchical co-scheduling scheme with task, operator, and data-level optimizations. 
Specifically, it fuses tasks through a hybrid of spatial and temporal multiplexing,
and orchestrates multi-task operator execution in two-tiered hybrid parallelism.
Additionally, \sysname{} employs chunk-based data alignment to mitigate inter-task ineffective tokens. 
Experimental results demonstrate that \sysname{} achieves up to $2.33\times$ higher throughput and $5.29\times$ memory reduction compared to three state-of-the-art baselines.

\end{abstract}

{
\renewcommand{\thefootnote}{}
\footnotetext[1]{$^*$Equal contribution.}
}

\pagestyle{empty}

\section{Introduction}

The paradigm of pretrain-finetune has surged in the realm of large language models (LLMs)~\cite{pretrain-finetune,bert,gpt-3,radford2019language,gpt-3,chrapek2024fortifyfoundationspracticalprivacy}. 
Enterprises and developers leverage domain-specific datasets to fine-tune a pretrained model into customized ones for more contextually relevant responses~\cite{rte,sst2,openbookqa,custom-llm-openai,custom-llm-penguin,codellama}.
With the evolution of emerging hardware and its rising cost, local fine-tuning has become increasingly prohibitive for many developers.
In this context, 
LLM service providers offer public \textit{fine-tuning APIs}~\cite{openai-peft,gemini-peft,togetherai-peft,open_pipe}, empowering developers to fine-tune LLMs remotely in shared GPU datacenters. 
Given that these APIs adopt a token-based pricing model~\cite{openai-peft,gemini-peft}, optimizing fine-tuning efficiency with minimal resources is crucial.

\begin{figure}
\centering
\includegraphics[width=.98\linewidth]{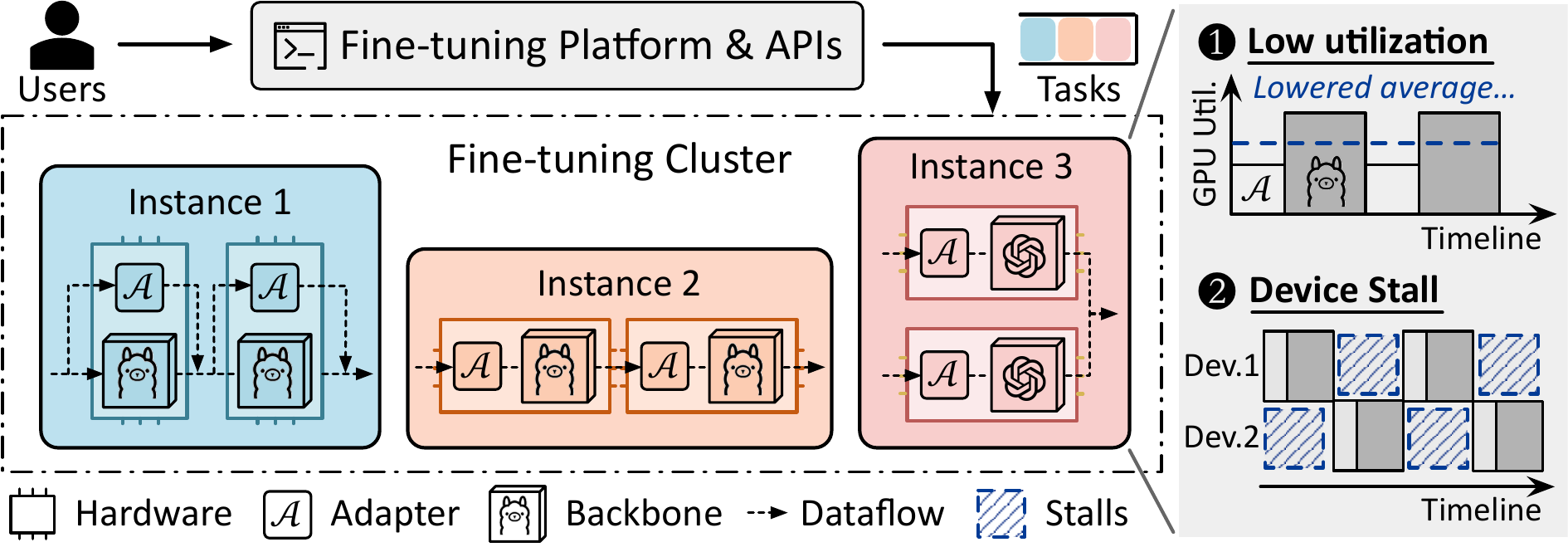}
\vspace{-1mm}
\caption{Workflow of submitting tasks and fine-tuning LLMs remotely. 
Each instance separately handles $1$ task of diverse PEFT types, with its backbone parallelized on $2$ GPUs.
}
\label{fig:workflow}
 \vspace{-4mm}
\end{figure}

To reduce service provisioning costs and accommodate diverse downstream domains, parameter-efficient fine-tuning (PEFT) is widely applied as one of 
the backends for these APIs~\cite{peft_survey,openai-peft,gemini-peft,open_pipe}.
Figure~\ref{fig:workflow} depicts an example workflow of fine-tuning services. 
A PEFT task attaches small-scale trainable \textit{adapters} to targeted operators of a pretrained LLM \textit{backbone}.
Only the adapters are fine-tuned, while the backbone parameters remain frozen~\cite{hift,lora}.
As illustrated, developers create PEFT tasks using fine-tuning APIs, while the cluster scheduler \textit{separately} dispatches tasks, deploys instances (hardware and the backbone), and launches PEFT programs.

However, existing PEFT frameworks (e.g., HuggingFace PEFT~\cite{hf_peft}, NeMo~\cite{nemo}) suffer from prominent resource inefficiencies, stemming from two major reasons.
First, PEFT inherently introduces small but non-negligible (in terms of latency) operators, such as LoRA down-projection~\cite{lora}, degrading average GPU utilization by up to $1.47\times$ (\ding{182} in Figure~\ref{fig:workflow}).
Some domain-specific PEFT corpora have shorter sequences and smaller batch sizes compared to pretraining~\cite{sentiment,short-text-gen}, further intensifying GPU underutilization.
Second, PEFT adapters exacerbate device stalls arising from communication delays and data dependencies between operators, in both intra-~\cite{megatron} and inter-stage~\cite{gpipe,pipedream} parallelism (\ding{183} in Figure~\ref{fig:workflow}).
Without backbone weight gradients, PEFT does not support stall-free pipeline, such as DeepSeek-V3 DualPipe~\cite{deepseekv3} and ZeroBubble ZB-H2~\cite{zerobubble}), which splits the backward pass into input and weight gradients for finer scheduling.
Communication overlapping techniques~\cite{transformer_engine,wang2022overlap}, which decompose computation operators into smaller tiles to overlap with communication, are prone to GPU underutilization and ultimately inflate end-to-end latency (\S\ref{sec:peft_issues}).
As GPU compute capability grows faster than interconnect bandwidth, the above inefficiencies are being further aggravated --- with a $3.18\times$ gap in Model FLOPs Utilization (MFU)~\cite{megascale} from NVIDIA H100 GPUs with NVLink to A40 GPUs with PCIe4.0 connection.
To improve GPU utilization and mitigate device stalls, in multi-tenant datacenters, we move beyond single-task optimization to multi-task co-scheduling by multiplexing the LLM backbone.
We propose to address the inefficiencies via \textit{spatial–temporal backbone multiplexing}, where a single backbone is shared across tasks by batching spatially and interleaving temporally.
While some multiplexing strategies are adopted in pretraining or serving~\cite{slora,nanoflow,chen2024centauri}, they have critical limitations when directly reused in PEFT scenarios.

Specifically, batching-based spatial multiplexing adapted from multi‑LoRA serving (e.g., SLoRA~\cite{slora}, Punica~\cite{punica}) can amplify device stalls and introduce ineffective computation, as PEFT workloads differ fundamentally from serving (\S\ref{sec:strawman}). 
On the other hand, temporal multiplexing and overlapping techniques, which are inspired by recent pretraining and serving systems (e.g., Centauri~\cite{chen2024centauri}, NanoFlow~\cite{nanoflow}), alleviate some communication bottleneck yet at the cost of even lower GPU utilization.
Direct spatial or temporal multiplexing approaches not only fail to efficiently resolve these resource inefficiencies but may also conflict with one another.

We therefore propose \textbf{\sysname{}}, a resource‑efficient system for multi-task PEFT that enables spatial-temporal backbone multiplexing in datacenters.
Nevertheless, several challenges arise across 
the system design and scheduling. 
At the task level, it is non-trivial to make the hybrid decision of spatial and temporal multiplexing across tasks to tame their inherent tradeoff.
At the operator level, efficiently orchestrating intricate operator execution across spatial and temporal multiplexing remains challenging under hybrid parallelism.
At the data level, efficiently aligning variable-length sequences from spatially batched tasks while mitigating inter-task ineffective tokens (i.e., zero padding) requires careful design.

The core idea of \sysname{} is unifying spatial and temporal multiplexing within a hierarchical design, built on top of unified PEFT representations, to optimize utilization and device stalls jointly.
This approach first modularizes diverse PEFT types into unified representations, enabling flexible sharing of the backbone with consistent training behavior (\S\ref{sec:backbone}).
Building on this foundation, \sysname{} incorporates a hierarchical multi-task co-scheduling scheme with tailored solutions at each level.
At the task level, \sysname{} introduces a \textit{hybrid task} (``hTask'') abstraction to navigate the spatial-temporal tradeoff: tasks within the same hTask are spatially batched to improve utilization, while different hTasks are temporally interleaved to hide stalls.
Guided by a pipeline-based cost model, it uses a dynamic programming (DP) algorithm to determine the optimal way to fuse tasks into multiple hTasks (\S\ref{sec:data}).
At the operator level, \sysname{} orchestrates operator execution of hTasks on the shared, modularized backbone.
By solving a two-tiered scheduling problem across inter- and intra-stage parallelism\footnote{Intra-stage refers to data parallelism (DP), tensor parallelism (TP)~\cite{megatron} and other variants~\cite{sequence_parallelism,gshard}, inter-stage refers to pipeline parallelism (PP)~\cite{pipedream}.}, it coordinates intricate dependencies under hybrid parallelism to generate a fine-grained, stall-free execution schedule (\S\ref{sec:orchestration}).
At the data level, \sysname{} aligns variable-length data across tasks within each spatially fused hTask via sequence packing and chunk-based partitioning (\S\ref{sec:alignment}).
The main contributions of this work are:

\begin{itemize}[itemsep=0pt, parsep=0pt, labelsep=5pt, leftmargin=*, topsep=3pt,partopsep=0pt]
\itemsep3pt
    \item We reveal the resource inefficiencies of PEFT workloads in datacenters, and identify the opportunities and challenges for efficient multi-task concurrent execution.

    \item We present \sysname{}, an efficient and scalable fine-tuning system for multi-task PEFT models via hierarchical spatial-temporal backbone multiplexing.

    \item We propose modularized backbone sharing for flexible and scalable spatial-temporal multiplexing. Building on this, we devise fine-grained, hierarchical co-scheduling scheme that integrates task, operator, and data-level optimizations to systematically optimize resource efficiency.
    
    
\end{itemize}

 We build \sysname{} and evaluate it with four LLMs and three PEFT datasets on testbeds with NVIDIA A40 and H100 GPUs. Empirical results show that \sysname{} achieves up to $2.33\times$ higher throughput and $5.29\times$ memory reduction compared to three state-of-the-art baselines. \sysname{} is open-sourced at \url{https://github.com/sjtu-epcc/muxtune}.
 
\section{Background and Motivation}

\subsection{Parameter-Efficient Fine-Tuning (PEFT) 
}\label{sec:peft_basic}

\begin{figure}
\centering
\includegraphics[width=.87\linewidth]{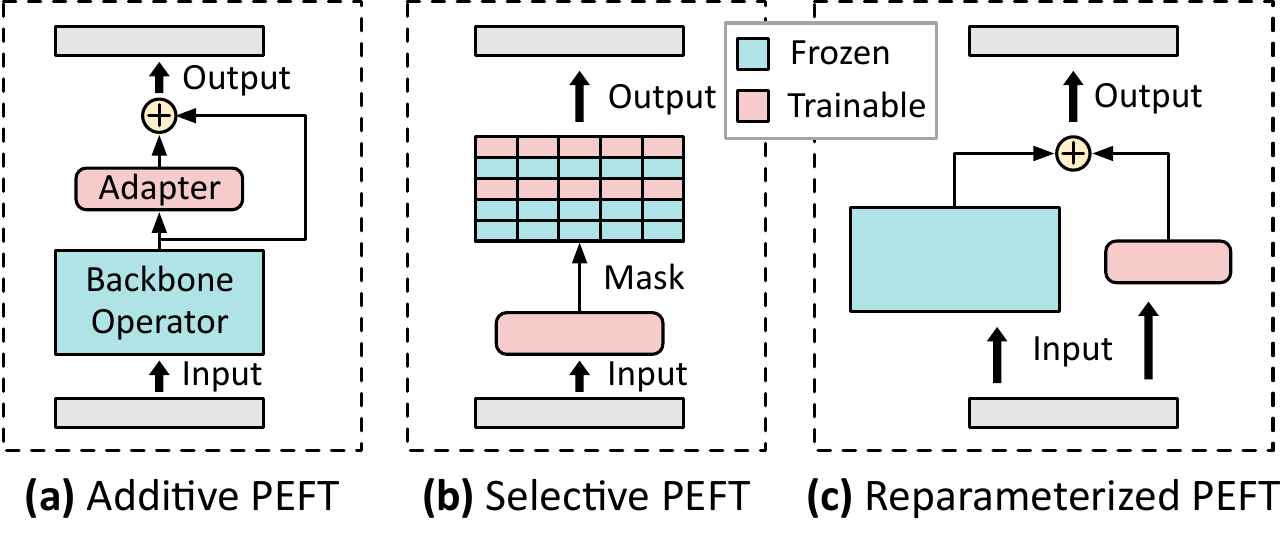}
\vspace{-2mm}
\caption{Representative categories of PEFT algorithms.
}
\label{fig:peft_types}
\vspace{-5mm}
\end{figure}

\paragraph{Basics.}
PEFT employs trainable adapters to adapt pretrained LLMs to specific domains cost-effectively, instead of training LLMs from scratch.
As shown in Figure~\ref{fig:peft_types}, it consists of three representative categories~\cite{peft_guide,peft_survey}:
(1) \textit{Additive} (e.g., Adapter-Tuning~\cite{adapter-tuning}) that inserts adapters into specific positions of the model architecture.
(2) \textit{Selective} (e.g., Diff-Pruning~\cite{diff-pruning}) that fine-tunes a subset of parameters via binary masks.
(3) \textit{Reparameterized} (e.g., LoRA~\cite{lora}) that constructs low-rank transformation of the original model parameters.

\paragraph{Characteristics.}
PEFT workloads in datacenters exhibit distinct characteristics. 
First, PEFT enables flexible configuration and attachment of adapters without affecting the backbone, as they manage independent parameters. 

Second, many PEFT tasks are fine-tuned on the same backbone type, as there are thousands of developers~\cite{openai_ft_user} yet a few available backbones (e.g., $7$ for OpenAI~\cite{openai-peft}, $3$ for Gemini~\cite{gemini-peft}).
These tasks retain independent and variable-length batches (i.e., \textit{data heterogeneity}), while sharing backbone operators except for adapter-related ones (i.e., \textit{backbone homogeneity}).
Some PEFT-based multimodal works have also proposed fine-tuning multiple adapters on a single backbone using the same corpus, each as an independent task~\cite{context-peft,adapter-fusion}.

Third, service providers have full access to model specifics, as they define, instantiate, and execute models when providing fine-tuning APIs.
This differs from conventional infrastructure providers, which only allocate hardware and run scripts, with tasks internally optimized by their frameworks.

Lastly, compared with pretraining, some domain-specific PEFT corpora, such as sentiment analysis~\cite{sentiment} and short text generation~\cite{short-text-gen}, feature shorter sequence lengths (e.g., 64) and smaller corpus sizes~\cite{smaller-dataset-arxiv,smaller-dataset}.
To ensure model generalization, prior works~\cite{smaller-bs-arxiv,lora} advocate adopting smaller batch sizes for model fine-tuning (e.g., $128$ for GPT-3~\cite{lora}).
Moreover, sequence lengths vary significantly across different PEFT corpora due to their varying domain focuses~\cite{sst2,openbookqa,rte}.

\subsection{Inefficiencies of PEFT Workloads}\label{sec:peft_issues}

\begin{figure}
\centering
\includegraphics[width=.98\linewidth]{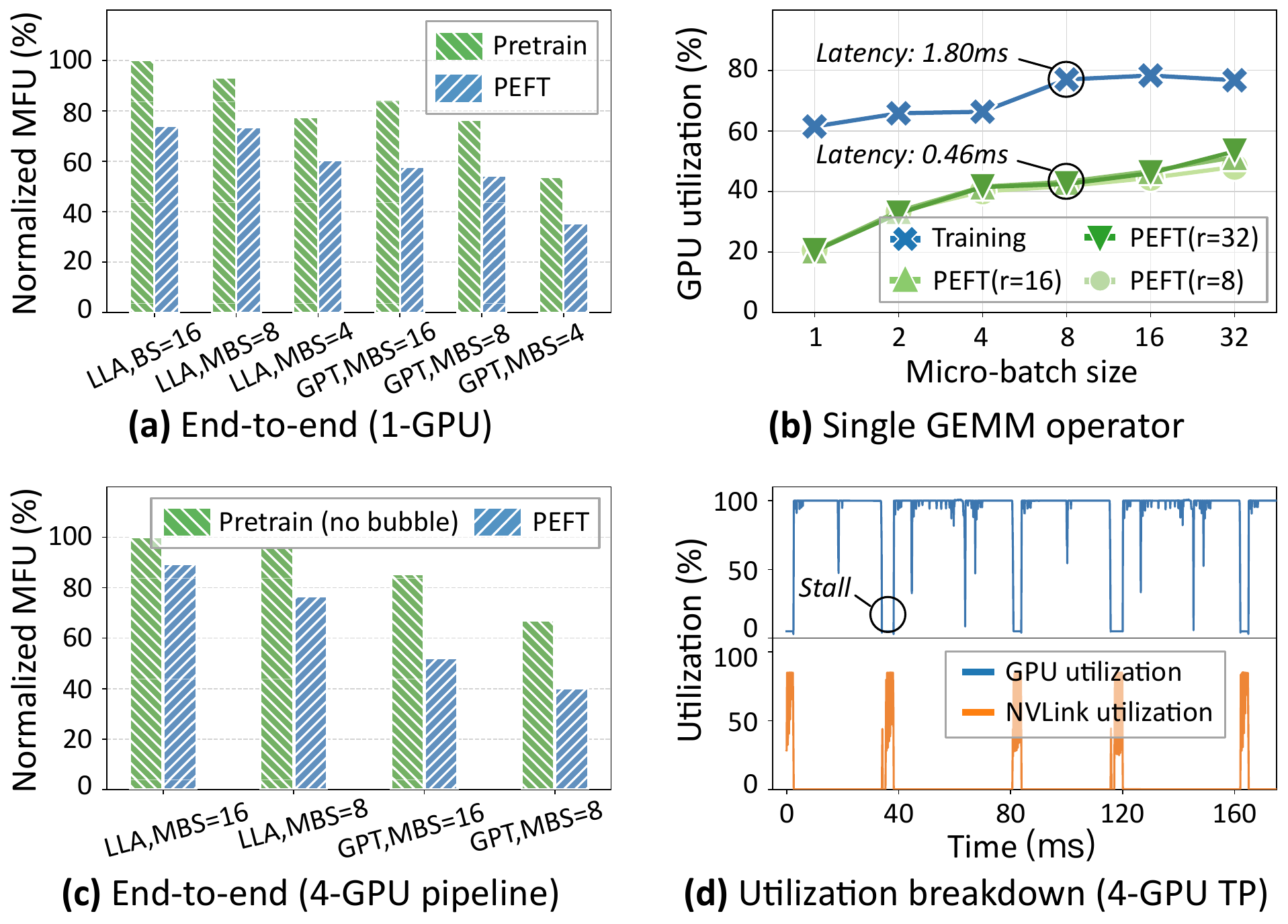}
\vspace{-1mm}
\caption{PEFT inefficiencies (\texttt{MBS}: micro-batch size, sequence length 128).
(a) Single-GPU MFU of $8$-layer models (global batch size $32$, \texttt{LLA}: LLaMA7B, \texttt{GPT}: GPT2.7B).
(b) Operator utilization (shape \texttt{[MBS,128,4096]$\times$[4096,r]}, \texttt{r=4096} for pretraining).
(c) Multi-GPU MFU of full models (global batch size $128$). 
(d) GPU and NVLink utilization.
}
\label{fig:peft_issues}
\vspace{-2mm}
\end{figure}

PEFT workloads typically suffer from resource inefficiencies on advanced GPUs (\S\ref{exp:setup}), as shown in Figure~\ref{fig:peft_issues} with two metrics: 
(1) \textit{GPU utilization}, which measures the occupancy of streaming multiprocessors (SMs)~\cite{ncu}; 
(2) \textit{Model FLOPs Utilization} (MFU), which measures end-to-end efficiency~\cite{bytecheckpoint}.

\paragraph{Insufficient Utilization.}
Figure~\ref{fig:peft_issues}(a) shows that PEFT exhibits lower single-GPU MFU than model pretraining (up to $1.47\times$).
The main reason is that PEFT omits compute-intensive backbone weight gradients and introduces small-scale operators, such as LoRA down-projection~\cite{lora} and learnable vectors of Prefix-Tuning~\cite{prefix-tuning}.
For example, LoRA rank (up to size 64) is $64.0\times$ smaller than the hidden size of LLaMA7B (size 4096).
Figure~\ref{fig:peft_issues}(b) shows that these small operators incur non-negligible latency (e.g., $0.46$ms vs. $1.80$ms in pretraining) and GPU underutilization (a gap of up to $40.9\%$).
Their attachment to non-continuous backbone operators (e.g., \texttt{qkv-proj} per decoder block~\cite{transformer}) prevents direct vertical or horizontal fusion due to stringent interdependencies.
Moreover, since PEFT corpora typically feature shorter sequences and smaller batch sizes (\S\ref{sec:peft_basic}), the input size remains limited and further undermines the efficiency of GPU parallel computing.
Worse yet, our experiments across GPU architectures (configured as Figure~\ref{fig:peft_issues}(a)) show that the average MFU of PEFT on NVIDIA V100, A40, and RTX6000 is $0.84\times$, $0.68\times$, $0.59\times$ that of pretraining, demonstrating that the underutilization is exacerbated by higher-end hardware.

\begin{figure}
\centering
\includegraphics[width=.98\linewidth]{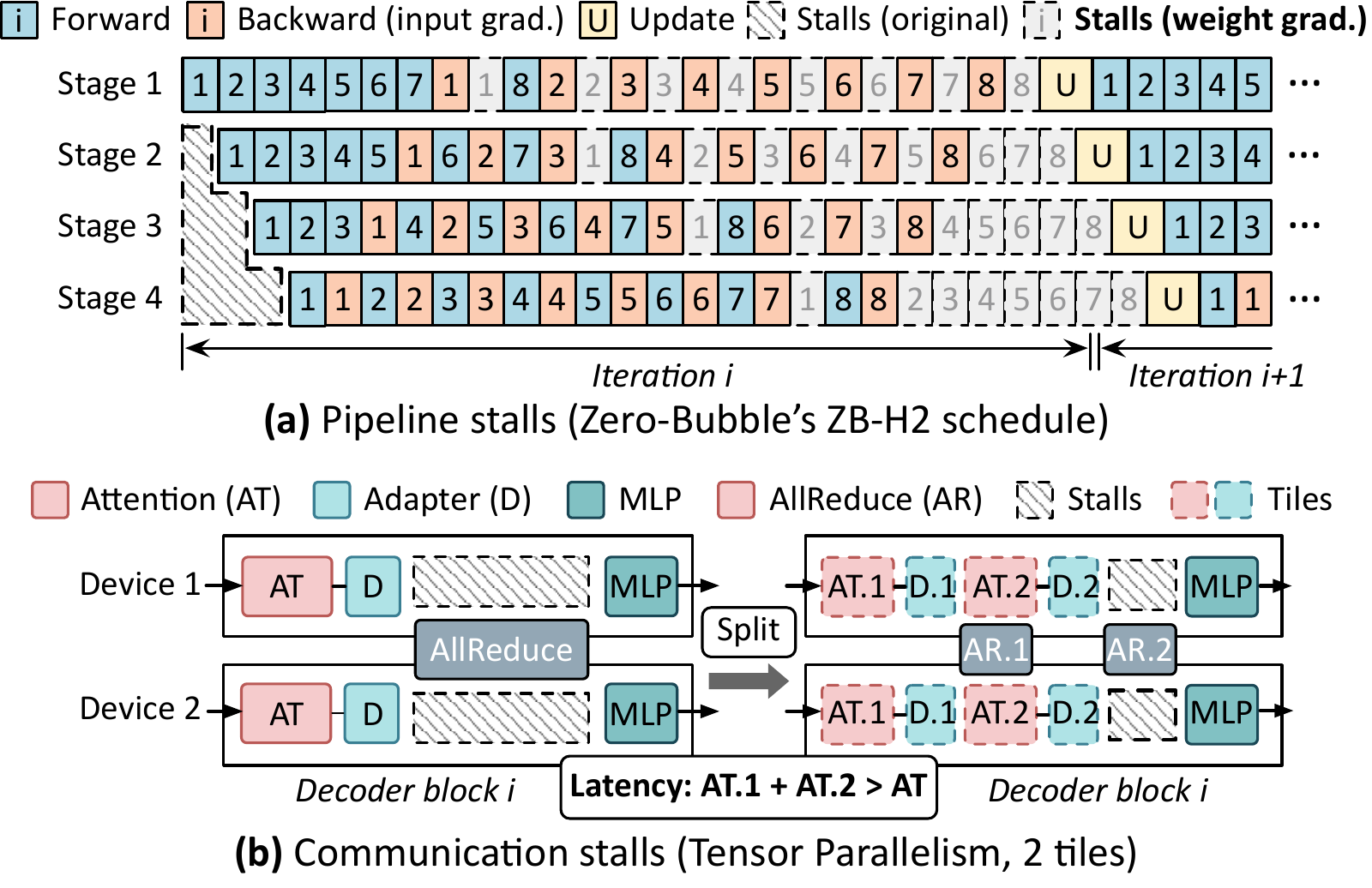}
\vspace{-2mm}
\caption{Device stalls in PEFT under model parallelism.
}
\label{fig:stalls}
\vspace{-4mm}
\end{figure}

\paragraph{Device Stall.}
In model parallelism, device stalls arise from communication delays and data dependencies between parallelized operators.
PEFT adapters exacerbate stalls by increasing stage latencies and intra-stage communication costs~\cite{slora}.
Worse still, advanced training optimizations, such as stall-free pipeline~\cite{deepseekv3,zerobubble} and communication overlapping~\cite{transformer_engine,wang2022overlap}, cannot be directly reused in PEFT. 
This leads to further MFU degradation during multi-GPU execution, with a drop of up to $1.65\times$ (Figure~\ref{fig:peft_issues}(c), worse than \texttt{1-GPU} case).
Figure~\ref{fig:peft_issues}(d) shows GPU and NVLink utilization in 4-GPU tensor parallelism with significant stalls.
Below, we demonstrate two stall types and discuss how PEFT aggravates them.


The first is \textit{pipeline stalls} (also known as bubbles) arising from pipeline flushes and inter-stage dependencies (Figure~\ref{fig:stalls}(a)).
Prior works (e.g., DualPipe~\cite{deepseekv3}, ZB-H2~\cite{zerobubble}) reduce bubbles by splitting the backward pass into input and weight gradients, finely scheduling for near-zero-bubble pipeline.
PEFT inherently lacks support for this technique due to the absence of backbone weight gradients.
As depicted, unlike original stalls, those stalls induced by omitted weight gradients grow linearly with the number of micro-batches, thus cannot be amortized.
Directly adopting DualPipe in PEFT undermines throughput by $1.16\times$ compared to 1F1B~\cite{pipedream}.

The second is \textit{communication stalls} in tensor parallelism caused by communication delays and operator dependencies (Figure~\ref{fig:stalls}(b)).
Prior works~\cite{transformer_engine,wang2022overlap} reduce these stalls by decomposing computation into smaller tiles to overlap with communication.
As shown, since PEFT inherently suffers underutilization, such decomposing intensifies the issue (a $24.5\%$ utilization drop) and inflates overall latency by $1.17\times$ for GPT2.7B with 2 GPUs.
Other stall types such as those from parameter synchronization~\cite{zero_offload,fsdp} also exist in PEFT.

\subsection{Intuitive Multiplexing Approaches}\label{sec:strawman}

\begin{figure}
\centering
\includegraphics[width=.98\linewidth]{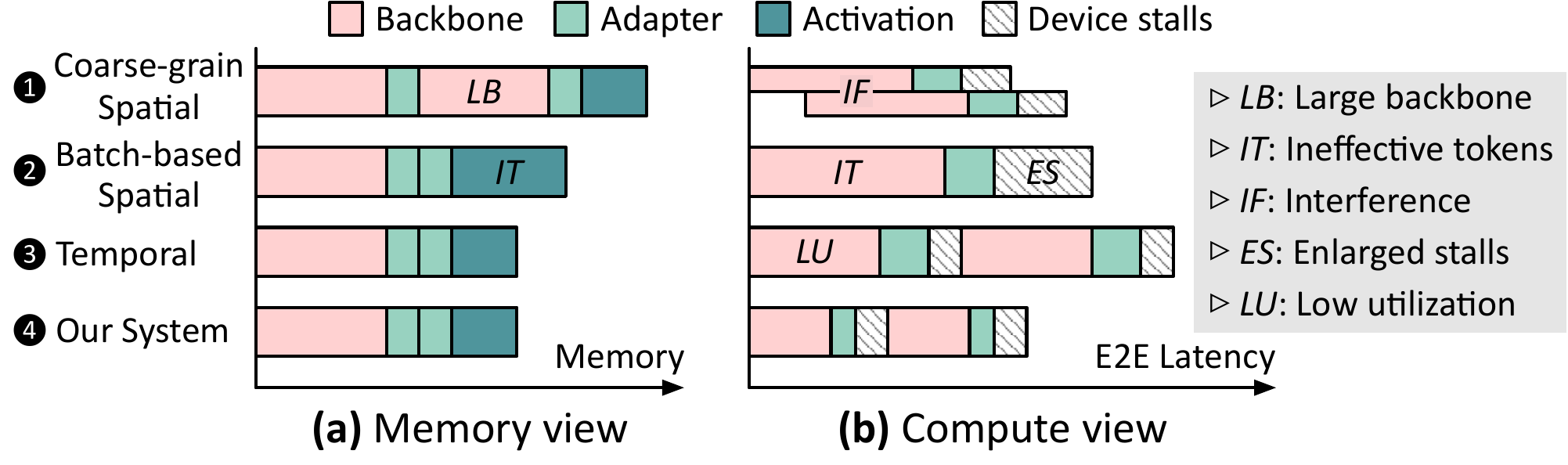}
\vspace{-1mm}
\caption{General views of intuitive multiplexing approaches.
}
\label{fig:strawman}
\vspace{-4mm}
\end{figure}

\paragraph{Coarse-grained spatial multiplexing causes poor scalability and limited performance gains.}
A straightforward way to improve utilization is co-locating PEFT tasks via NVIDIA MPS~\cite{mps} or multiple streams~\cite{cuda_stream}, similar to~\cite{gavel,lucid}.
However, the utility of this approach is limited by memory constraints (\ding{182} in Figure~\ref{fig:strawman}).
To illustrate, we profile the memory breakdown of a LoRA LLaMA7B (batch size $8$, sequence length $128$): the backbone parameters and activations consume $13.4$GB and $4.3$GB, respectively, while the total memory footprint is $18.1$GB. 
With 4 A40 GPUs (48GB each), only $8$ tasks can be co-located without parallelization,
preventing scaling to larger models or more concurrent tasks.
Additionally, the lack of fine-grained control over operator execution results in suboptimal stall reduction and potential inter-task interference (e.g., a $2.5\times$ performance drop reported in~\cite{nanoflow}).

\paragraph{Batching-based spatial multiplexing causes enlarged stalls and ineffective computation.}
Prior multi-LoRA serving systems~\cite{slora,punica} simplistically batch multiple requests for concurrent computation.
While effective for serving, this approach is misaligned with PEFT workloads (\ding{183} in Figure~\ref{fig:strawman}).
First, backward pass introduces pipeline stall concerns that are generally absent in forward-only serving~\cite{deepspeed-inference}.
Excessive batching exacerbates device stalls and undermines end-to-end performance (\S\ref{sec:data}).
Second, batching offers diminishing returns in compute-bound PEFT, compared to the memory-bound decoding phase of serving~\cite{sarathi}.
Lastly, continuous batching~\cite{orca} used in serving avoids the need for padding, which remains essential yet computationally ineffective to align variable-length sequences in PEFT (\S\ref{sec:alignment}).

\paragraph{Temporal multiplexing and overlapping result in even lower GPU utilization.}
The approach of temporally overlapping communication with computation from other tasks, inspired by recent pretraining and serving systems~\cite{chen2024centauri,nanoflow}, also proves counterproductive for PEFT (\ding{184} in Figure~\ref{fig:strawman}).
Specially, temporal multiplexing executes operators of each task sequentially, instead of exploiting batching opportunities to improve intra-operator utilization.
Given the limited input sizes typical of PEFT workloads, it may also be impractical to excessively increase the micro-batch size for a single task.


\subsection{Our Approach and Challenges}\label{sec:challenges}

We identify an opportunity to address the limitations via spatial–temporal backbone multiplexing, i.e., sharing the backbone across tasks by batching spatially and interleaving temporally. 
This paradigm forms an intricate optimization space with multiple \textit{coupled} dimensions, e.g., determining the optimal spatial-temporal combination while scheduling multi-task pipeline execution. 
To reduce optimization complexity, we propose \textit{hierarchically decomposing} the space into three levels and tackling the unique challenges at each level:


\textbf{Task Level: Navigating spatial-temporal tradeoff.}
There is an inherent tradeoff between spatial and temporal multiplexing (\S\ref{sec:data}).
This poses a primary challenge to design a dynamic scheduling policy that can intelligently batch tasks spatially to improve utilization while interleaving them temporally to hide the pipeline and communication stalls.

\textbf{Operator Level: Stall-free multi-task orchestration. }
Coarse-grained spatial multiplexing fails to mitigate stalls without operator-level execution control.
Under hybrid parallelism, operator orchestration across spatially and temporally multiplexed tasks becomes a complicated two-tiered problem (\S\ref{sec:orchestration}). It requires tailored algorithms and fine-grained coordination for stall-free execution and high GPU utilization.

\textbf{Data Level: Mitigating inter-task ineffective computation.}
Spatially multiplexing tasks with variable-length sequences naturally requires aligning along sequence dimension.
Na\"ive strategies such as zero padding or packing into long sequences either waste compute on ineffective tokens or degrade efficiency.
Without careful alignment design, these effects can diminish the gains from multi-task multiplexing.


\section{System Design}

\subsection{Overview}\label{sec:overview}

\begin{figure}
\centering
\includegraphics[width=.9\linewidth]{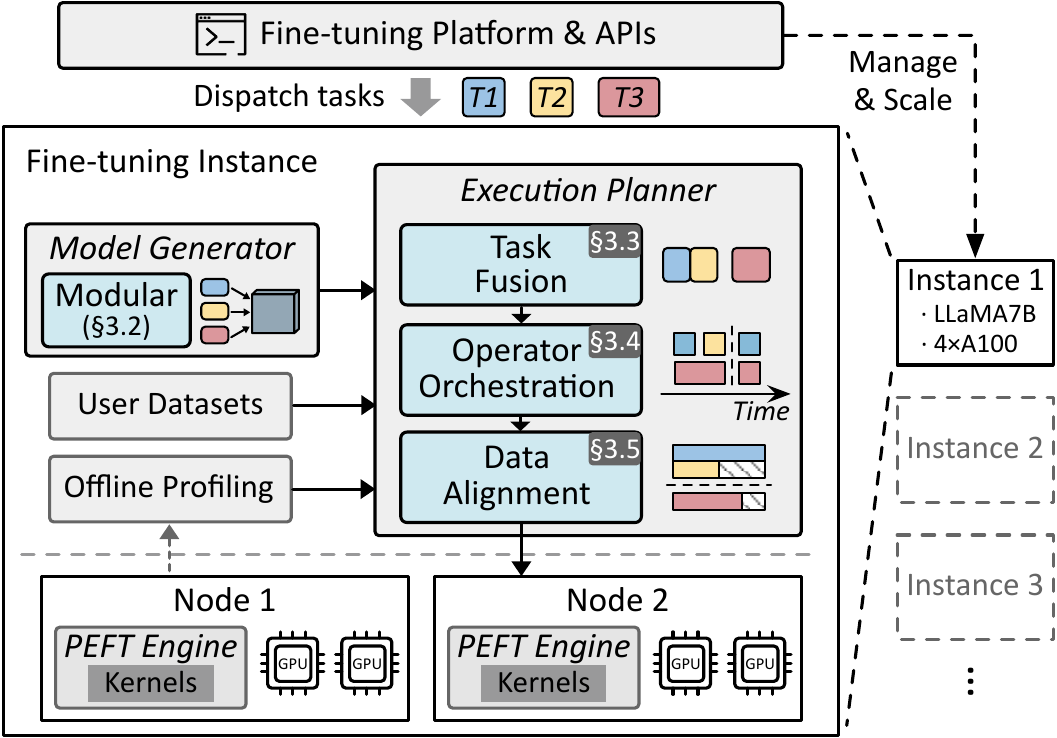}
\vspace{-1mm}
\caption{Architecture overview of \sysname{}.
}
\label{fig:overview}
\vspace{-5mm}
\end{figure}

\sysname{} is an efficient and scalable system for multi-task PEFT, serving as the backend for LLM fine-tuning APIs to enhance resource efficiency in multi-tenant GPU datacenters.
The key idea of \sysname{} is to multiplex the backbone in a spatial-temporal manner across independent PEFT tasks via flexible, modularized backbone sharing and fine-grained, hierarchical multi-task co-scheduling.

Figure~\ref{fig:overview} presents the architecture of \sysname{} with three main modules: model generator, planner, and PEFT engine to address the challenges in \S\ref{sec:challenges}.
Initially, users configure their PEFT tasks (e.g., backbone type) and submit tasks via fine-tuning APIs. The cluster scheduler dispatches tasks with the same backbone to an in-flight instance or creates a new one based on scheduling policies (e.g., budget-based~\cite{kube}).
Given dispatched tasks, \textit{Model Generator} builds a PEFT model with multi-task adapters based on modularization (\S\ref{sec:backbone}).
With the PEFT model, user datasets, and profiling data, \textit{Execution Planner} fuses tasks into hybrid tasks by adaptively combining spatial and temporal multiplexing (\S\ref{sec:data}). 
Then, the planner orchestrates the fine-grained operator execution of these hybrid tasks under two-tiered hybrid parallelism (\S\ref{sec:orchestration}).
From the data perspective, data batches are loaded in a streaming manner and aligned across spatially batched tasks (\S\ref{sec:alignment}).
At runtime, \textit{PEFT Engine} concurrently executes the dispatched PEFT tasks via efficient fused and overlapped kernels.

\subsection{Backbone Sharing}\label{sec:backbone}

\sysname{} adopts flexible, modularized backbone sharing to co-locate PEFT tasks with diverse workloads.
This subsection introduces PEFT modularization and sharing the backbone across multi-task sub-modules for efficient multiplexing.

\begin{figure}
\centering
\includegraphics[width=\linewidth]{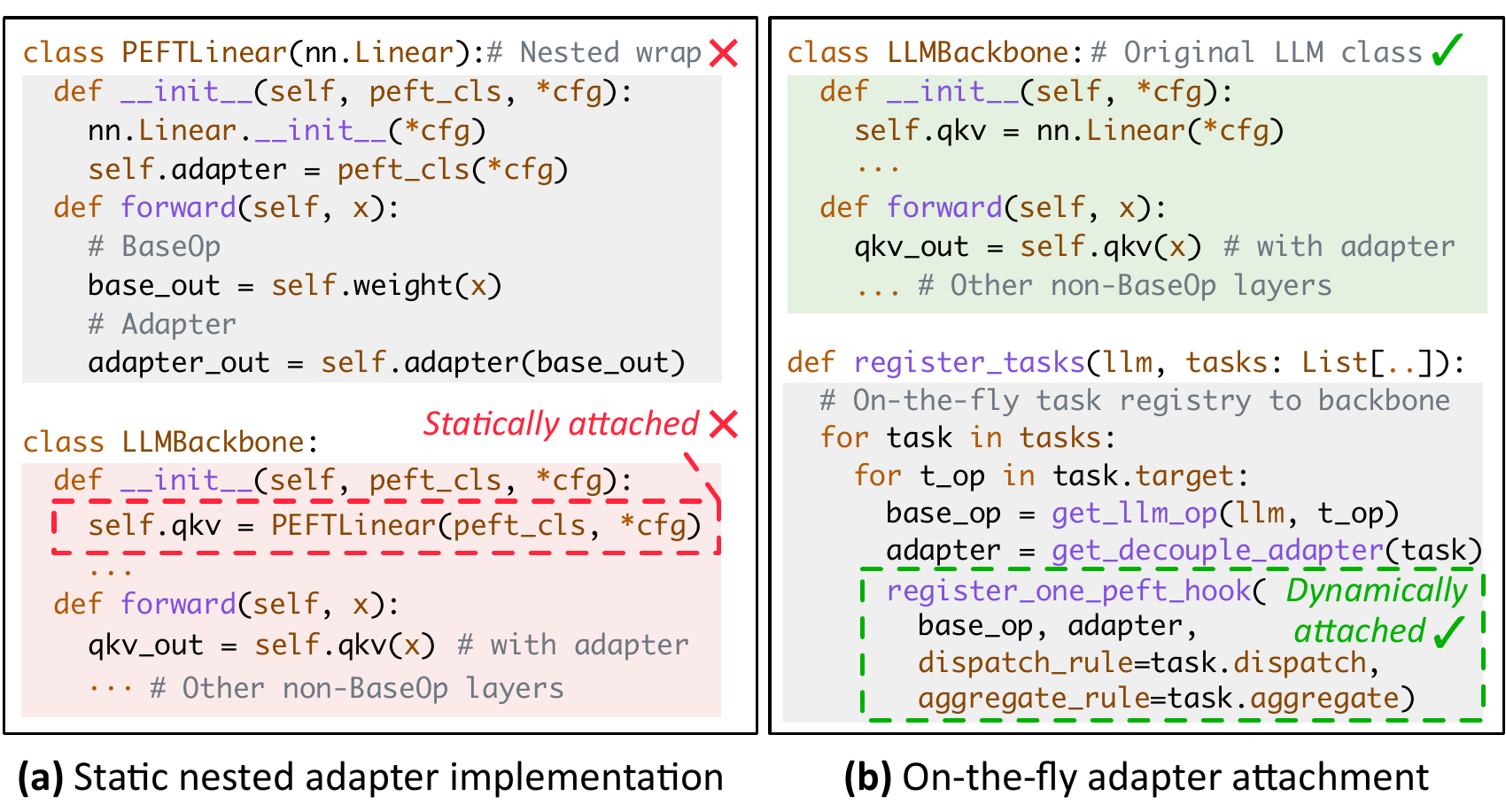}
\vspace{-6mm}
\caption{Current static nested adapter implementation and our modularization-based dynamic adapter attachment.
}
\label{fig:modular}
\vspace{-4mm}
\end{figure}

\paragraph{PEFT Modularization.}
As shown in Figure~\ref{fig:modular}(a), existing single-task frameworks~\cite{hf_peft,nemo} \textit{statically} inject adapters into the LLM backbone, i.e., directly treating adapters of a task as LLM layers.
While intuitive, employing such a static implementation in multi-task PEFT scenarios cannot support dynamic workloads with shifting adapter numbers and types, as it requires from-scratch model reinitialization to tackle on-the-fly task arrival or completion events.

In contrast, \sysname{} employs \textit{decoupled adapters} to support flexible user customization and efficient scaling by the cluster scheduler.
Specifically, by examining mainstream PEFT types in \S\ref{sec:peft_basic}, we abstract the general PEFT workflow into four sub-modules:
(\romannumeral1) \textit{BaseOp} is an operator of the backbone that has the possibility to be attached an adapter, such as QKV and Linear Projection (Attention is excluded).
(\romannumeral2) \textit{Adapter} describes the PEFT algorithms (e.g., LoRA~\cite{lora}) and is customized by users.  
(\romannumeral3) \textit{Dispatch} defines the multi-task data dispatching rules to prepare input tensors for \textit{BaseOp} and \textit{Adapter}.
(\romannumeral4) \textit{Aggregate} defines the data aggregation rules to gather output tensors from \textit{BaseOp} and \textit{Adapter}.






\paragraph{Dynamic Multi-Task Backbone Sharing.}
As shown in Figure~\ref{fig:modular}(b), unlike prior frameworks, \sysname{} preserves non-intrusiveness to the LLM backbone while providing a \texttt{register\_tasks()} API to handle on-the-fly task arrival events.
This is the cornerstone of multi-task backbone sharing.
When the cluster scheduler assigns new tasks to an in-flight instance, the local model generator reactively invokes this API to register on the multiplexed backbone without costly model reinitialization (\S\ref{sec:implement}). 
After that, \sysname{} implicitly automates the task, operator, and data-level optimizations as introduced below, and efficiently executes the generated multi-task PEFT model at runtime.


\paragraph{Isolation and Convergence Guarantee.}
While improving efficiency via multi-task backbone sharing, \sysname{} provides stringent guarantee for both inter-task isolation and consistent model convergence.
Specifically, \sysname{} safely instantiates the LLM backbone and user-defined adapters (\S\ref{sec:peft_basic}), thereby preventing most runtime errors (e.g., semantic errors) described in~\cite{jeon2019analysis}.
Through fine-grained memory modeling and operator orchestration, \sysname{} also mitigates inter-task performance interference and OOM risks.

In \sysname{}, both \textit{BaseOp}s and \textit{Adapter}s across temporally interleaved tasks are naturally isolated in time, with no impact on convergence.
For spatially batched tasks, \textit{BaseOp}s are concatenated along batch dimension and exhibit mathematical isolation.
Taking a GEMM \textit{BaseOp} with two \textit{Adapter}s as an example, the forward computation of \textit{BaseOp} is:
\begin{equation}
\begin{aligned}
BW_{g} = [B_1, B_2]_{b} W_{g} = [B_1W_{g}, B_2W_{g}]_{b}, \quad \text{(\textit{BaseOp} fwd)}
\label{eq:base_fwd}
\end{aligned}
\end{equation}
where $B_{i}$ is the batch of task $i$, $B$ is the concatenated batch, and $W_{g}$ is GEMM weight.
In backward pass, each $B_i$ is independently computed a loss and backpropagated, while computations of the same \textit{BaseOp} are batched as:
\begin{equation}
\begin{aligned}
G^{in} = [G^{in}_1, G^{in}_2]_b \leftarrow [G^{out}_1, G^{out}_2]_b W^\mathrm{T}_g, \quad \text{(\textit{BaseOp} bwd)}
\label{eq:base_fwd}
\end{aligned}
\end{equation}
where $G^{in}_i$ and $G^{out}_i$ are the input and output gradient of task $i$. $G^{in}$ is the transient gradient buffer of the input to \textit{BaseOp}.
Other non-batchable operators (e.g., adapters) are independently computed in a fused manner (\S\ref{sec:adapter_fusion}).
Such isolation preserves consistent convergence (e.g., $0.07$ mean-square deviation) while avoiding numerical failure propagation (e.g., gradient NaN from overlarge learning rate) among tasks.

\subsection{Task Fusion}\label{sec:data}

Based on modularized representation and flexible interface, \sysname{} is able to multiplex the backbone in different manners: (1) batching PEFT tasks spatially by executing batched \textit{BaseOp}, \textit{Dispatch}, fused \textit{Adapter} and \textit{Aggregate}; or (2) interleaving tasks temporally by sequentially executing \textit{BaseOp}, \textit{Dispatch}, \textit{Adapter}, and \textit{Aggregate} of each task.

\begin{figure}
\centering
\includegraphics[width=.98\linewidth]{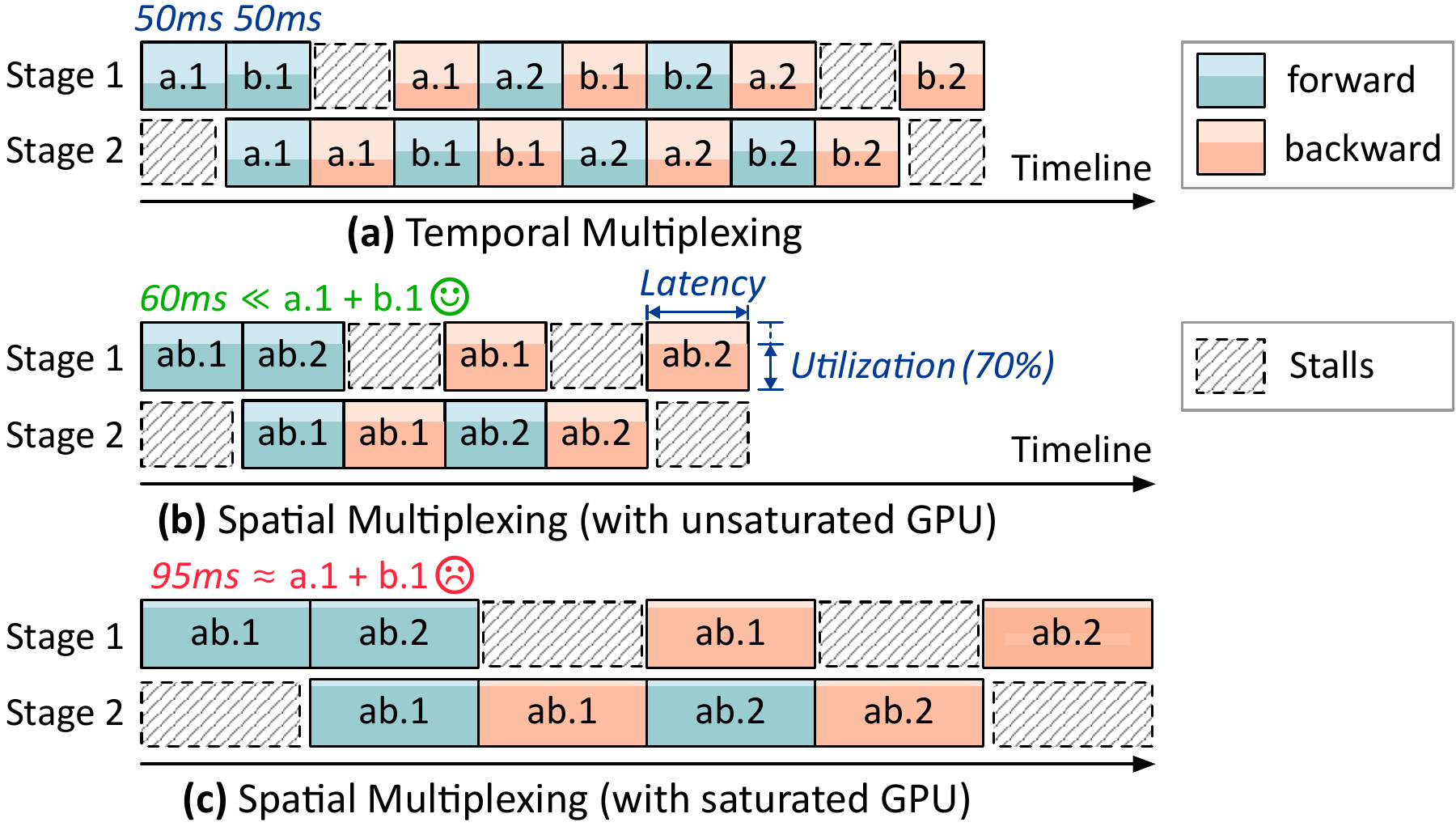}
\vspace{-1mm}
\caption{Tradeoff of spatial and temporal multiplexing. \texttt{x.i} represents $i$-th micro-batch of (hybrid) task \texttt{x}.
}
\label{fig:coalescing}
\vspace{-4mm}
\end{figure}

\paragraph{Spatial-Temporal Tradeoff.}
Figure~\ref{fig:coalescing} illustrates these two multiplexing options.
The \textit{interleaving-based temporal multiplexing} computes tasks sequentially, overlapping stalls with computations across tasks, but risks GPU underutilization with separate \textit{Adapters} and limited input sizes~(\S\ref{sec:peft_issues}).
The \textit{batching-based spatial multiplexing} batches the \textit{BaseOp} and fuses the \textit{Adapter} of independent tasks to improve utilization. 
However, excessive batching prolongs operator (or stage) latency and exacerbates stalls in parallelized execution. 

\begin{figure}
\centering
\includegraphics[width=.98\linewidth]{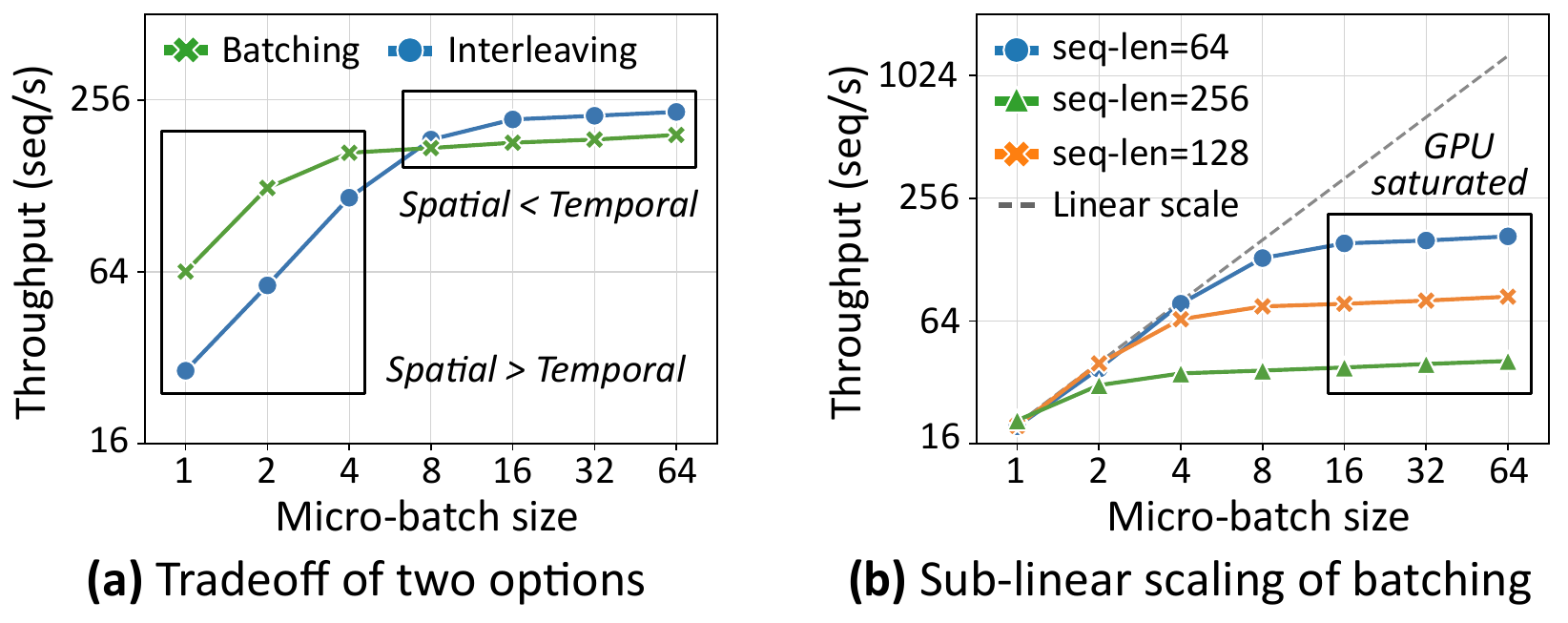}
\vspace{-2mm}
\caption{Quantified analysis of two options. (a) 2 tasks on 16-layer LLaMA7B with 4-GPU pipeline (4 micro-batches, \texttt{seq-len} 64). (b) 1 task on 8-layer LLaMA7B with 1 GPU.
}
\label{fig:tradeoff}
\vspace{-4mm}
\end{figure}

The optimal multiplexing decision depends on task input sizes, PEFT parameters, and parallelism strategy.
As shown in Figure~\ref{fig:tradeoff}(a), spatially batching tasks yields better performance when GPUs are unsaturated, while temporally interleaving is preferable at higher GPU utilization.
This shift stems from the diminishing returns induced by excessive spatial batching beyond GPU saturation (Figure~\ref{fig:tradeoff}(b)).
For example, ideally batching 8 tasks, each with micro-batch size 8 and sequence length 128, only improves throughput by $1.12\times$, far short of the expected $8\times$ gain.
Notably, the above analysis only involves two tasks, while scaling to more tasks exponentially elevates the complexity of such multiplexing decisions.

\paragraph{Hybrid Task Abstraction.}
Given the scheduling complexity, deriving the optimal multiplexing decision requires a hierarchical scheduling of all tasks.
We introduce a \textit{hybrid task} abstraction (``hTask'' for short) to unify the two multiplexing options, which is fused from independent tasks.
Within a hTask, tasks are fused and batched in a spatial manner; among hTasks, their computations are temporally interleaved.

The task fusion is framed as a bin-packing problem.
Initially, $M$ tasks $\mathcal{T} = \{\mathcal{T}_1, \mathcal{T}_2, ..., \mathcal{T}_M\}$ co-locates on the backbone ($S$ stages, each has $N^{(s)}_{g}$ GPUs). 
As latency correlates positively with the input size (due to the backbone homogeneity described in \S\ref{sec:peft_basic}), we sort $\mathcal{T}$ by token count (denoted as $n_i$ for the $i$-th task) in ascending order.
Let $\mathcal{H}_{i\rightarrow j}$ be the hTask with task $i$ to $j$.
A unified number of micro-batches $C$ for all tasks is set empirically.
Sequences of each task are padded to a maximum length $l_i$, as detailed in \S\ref{sec:alignment}.

\paragraph{Cost Model.}
We build a cost model to assess end-to-end latency and memory footprint under hybrid parallelism.
Since forward and backward passes of the same stage share similar latency in PEFT (due to the absence of weight gradients), we model the latency of a hTask $\mathcal{H}_{i\to j}$ for the $s$-th stage (\textit{St.}) as:
\begin{equation}
\begin{aligned}
{L}^{(s)}(\mathcal{H}_{i\to j})=\sum_{o \in \textit{St.}^{(s)}} t_o\left(\sum_{k} n_k\right) / N^{(s)}_g \quad \text{(\textit{BaseOp} Lat.)} \\
+\sum_{\{a\} \in \textit{St.}^{(s)}}\max\left\{
\begin{array}{l}
\sum_{k}u_a\cdot t_a(n_k), \\ \max_{k} t_a(n_k)
\end{array}\right\}, \quad \text{(\textit{Adapter} Lat.)}
\label{eq:model-latency}
\end{aligned}
\end{equation}
where the first line models the latency of \textit{BaseOp}s sharded across $N^{(s)}_g$ GPUs ($k\in [i, j]$).
The $t_o(x)$ is the latency of computation operator $o$ with $x$ tokens (communication is overlapped as in \S\ref{sec:intra-stage}).
The second line estimates the latency of fused adapters $\{a\}$, where $u_a(x)$ and $t_a(x)$ are the GPU utilization and latency of \textit{Adapter} $a$ with $x$ tokens.
We use the weighted sum $u_a\cdot t_a(x)$ to estimate the overall latency after horizontal fusion (\S\ref{sec:adapter_fusion}), while bounding it with the maximum per-adapter latency to avoid the bottleneck effect.
In the pipeline with $S$ stages, the end-to-end latency of $\mathcal{H}_{i\to j}$ is: 
\begin{equation}
\begin{aligned}
{L}(\mathcal{H}_{i\to j}) = 2 \sum_{s=1}^{S-1} \{{L}^{(s)}(\mathcal{H}_{i\to j})\} + 2C \max_{1\leq s \leq S} {L}^{(s)}(\mathcal{H}_{i\to j}),
\label{eq:pipeline-latency}
\end{aligned}
\end{equation}
where the first term estimates the latency sum of warm-up and drain phases, while the second term models the overall latency of the steady phase with $C$ micro-batches~\cite{dynapipe}.

As for memory footprint, we find that it mainly consists of three parts in PEFT: (\romannumeral1) backbone parameters ${M}_b$, (\romannumeral2) input gradients ${M}^{(i)}_g$ of task $i$, and (\romannumeral3) activation ${M}^{(i)}_a$.
For 1F1B pipeline, the maximum per-stage memory is estimated as: 
\begin{equation}
{M_{stage}} = [{M}_b + \sum_{i=1}^M {M}^{(i)}_g]/S + \sum_{i=1}^M {M}^{(i)}_a(b_i, l_i), 
\label{eq:memory}
\end{equation}
where the first two terms are irrelevant to the input size (e.g., micro-batch size $b_i$). The third term is accumulated to at most $S$ copies and is proportional to $b_i$ and the sequence length $l_i$.
In practice, ${M}^{}_g$ typically resues the allocated memory of ${M}^{}_a$~\cite{pytorch}.
This memory model is used to assess whether a hybrid task (hTask) would cause Out-of-Memory (OOM) issues during the construction.
Evaluated in \S\ref{exp:ablation}, it precisely matches the scaling of the measured memory footprint.

\paragraph{Task Fusion with DP Algorithm.}
Based on cost modeling, we employ a dynamic programming (DP) algorithm to minimize the end-to-end pipeline latency when bin-packing $M$ tasks into $N$ hTasks.
We derive the state transition equation, where ${F}(m, n)$ is the minimal end-to-end latency of packing the first $m$ tasks into $n$ hTasks ($m \in [1, M]$, $n \in [1, N]$, $m \geq n$):
\begin{equation}
\begin{gathered}
{F}(m, n) = \min\limits_{n-1 \leq i \leq m} \{ {F}(i, n - 1) + L(\mathcal{H}_{(i+1)\to k}) / S \}, \\
{F}(m', 1) = L(\mathcal{H}_{1\to m'}), \quad \forall m' \in [1, M],
\end{gathered}
\label{eq:dp}
\end{equation}
where the impact of $\mathcal{H}_{(i+1)\to k}$ over ${F}(m, n)$ is estimated via average per-stage latency to satisfy the optimal substructure of DP.
This is because the steady phase typically dominates end-to-end latency, in which $\mathcal{H}_{(i+1)\to k}$ adds one forward-backward pass~\cite{terapipe}.
The optimal fusion plan is derived as $F^* = \min_{1 \leq N \leq M} \{ F(M, N) \}$.
Notably, minimizing end-to-end latency also balances the loads across hybrid tasks, as the pipeline is bottlenecked by its slowest stage~\cite{computer_arch}.
While the time complexity is $O(M^2(S + M))$, the algorithm incurs minimal overhead with modest number of tasks per backbone (\S\ref{exp:ablation}) and can be accelerated via parallel computing of $N$.


\subsection{Operator Orchestration}\label{sec:orchestration}

After task fusion, \sysname{} orchestrates hybrid task execution at the operator\footnote{To unify intra- and inter-stage parallelism, we use ``operator'' to represent both layer (e.g., GEMM) and stage-granularity computations.} granularity across spatial and temporal multiplexing in both intra- and inter-stage parallelism.

\begin{figure}
\centering
\includegraphics[width=\linewidth]{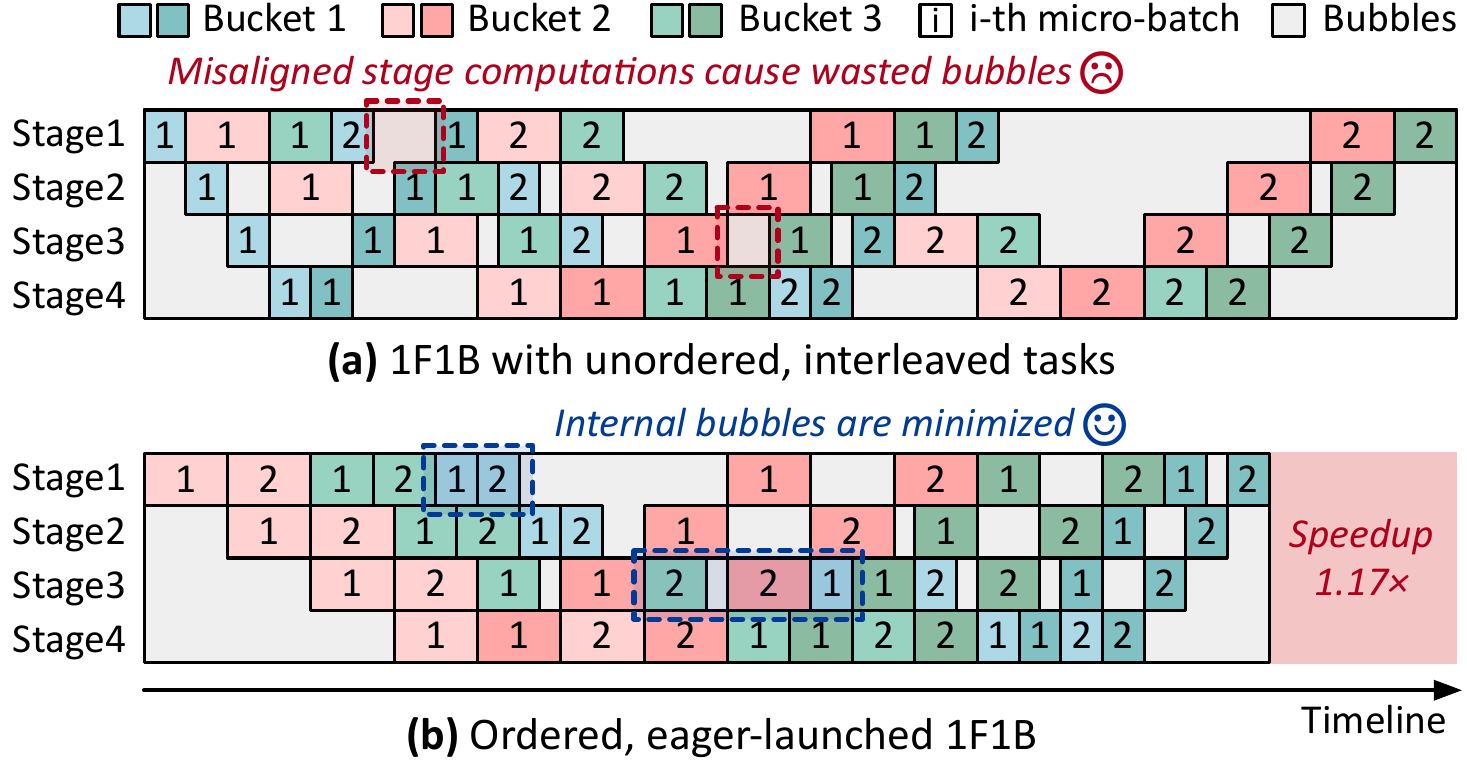}
\vspace{-6mm}
\caption{Example of inter-stage orchestration. Light/dark color represents forward/backward pass of hTask buckets.
}
\label{fig:inter_stage_orch}
\vspace{-3mm}
\end{figure}

\paragraph{Disaggregating Two-Tiered Orchestration.}
Given the distinct characteristics of intra- and inter-stage parallelism, we disaggregate operator orchestration by grouping $N$ hTasks $\mathcal{H}=\{\mathcal{H}_1, \mathcal{H}_2...,\mathcal{H}_N\}$ into $P$ buckets $\mathcal{G}=\{\mathcal{G}_1,\mathcal{G}_2,...,\mathcal{G}_P\}$.
Those hTasks of the same bucket are interleaved within a pipeline clock (i.e., intra-stage parallelism), while different buckets are interleaved across pipeline clocks (Figure~\ref{fig:inter_stage_orch}).





\paragraph{Workload-Balanced Hybrid Task Grouping.}
To efficiently determine the optimal grouping strategy, we decouple hTask grouping from operator orchestration, based on the observation that \textit{given a fixed $P$, balanced workloads lead to fewer internal bubbles and lower end-to-end latency}.
We traverse $P$ from $1$ (grouping all hTasks) to $N$ (one hTask in each bucket), optimizing $\mathcal{G}$ by minimizing inter-bucket variance: 
\begin{equation}
\begin{aligned}
\mathcal{G}^*(P) = &\mathop{\arg\min}_{\mathcal{G}=\{\mathcal{G}_1,...,\mathcal{G}_P\}} \sum_{j=1}^P |L^{(1)}(\mathcal{G}_j) - \overline{L^{(1)}(\mathcal{G})}|^2, \\
\mathrm{s.t.} \quad &L^{(1)}(\mathcal{G}_j) = \sum_{i}L^{(1)}(\mathcal{H}_i), \ \forall \mathcal{H}_i \in \mathcal{G}_j,
\label{eq:set_dev}
\end{aligned}
\end{equation}
where $L^{(1)}(\mathcal{H}_i)$ is the first stage latency of hTask $\mathcal{H}_i$, which is used because this step focuses on balancing workloads across groups. 
We then invoke the orchestration process below to model the end-to-end latency for each fixed $P$, and select the optimal grouping strategy $\mathcal{G}^*$ with the minimal latency.


\subsubsection{Inter-Stage Orchestration}\label{sec:inter-stage}
To optimize inter-stage execution with hTask buckets $\mathcal{G}=\{\mathcal{G}_1,...,\mathcal{G}_P\}$, the goal is to schedule the micro-batches of $\{\mathcal{G}_j\}_{1\leq j \leq P}$ to minimize end-to-end pipeline latency.


\paragraph{Computation Homogeneity.}
Micro-batch scheduling involves a re-entrant flow shop problem~\cite{graves1983scheduling}, where $\{\mathcal{G}_j\}_{1\leq j \leq P}$ are repeatedly processed by stages in forward and backward passes.
Unlike prior training works with complex scheduling~\cite{dynapipe}, we observe \textit{computation homogeneity} in PEFT:
(\romannumeral1) in each bucket $\mathcal{G}_j$, micro-batches retain a consistent shape across iterations (\S\ref{sec:alignment}), while shapes may vary across buckets;
(\romannumeral2) for bucket $\mathcal{G}_j$, forward and backward passes share the same execution time due to the absence of weight gradients.

These features enable us to prune the scheduling space and reduce algorithm overhead.
Specifically, with (\romannumeral1), \sysname{} adopts a unified pipeline template to execute multi-task iterations in a \textit{structured} manner, rather than relying on costly real-time scheduling per iteration. 
With (\romannumeral2), as shown in Figure~\ref{fig:inter_stage_orch}, \sysname{} efficiently minimizes internal bubbles by interleaving forward and backward passes of those buckets with similar latencies (e.g., the lower blue block of Figure~\ref{fig:inter_stage_orch}(b)).

\paragraph{Structured Pipeline Template.}
The pipeline template is dynamically generated based on bucket information, such as stage latencies and memory footprint, for structured multi-task execution.
Specifically, to mitigate bubbles, the template generation extends the 1F1B pipeline with three rules:
(1) Sorting $\mathcal{G}$ by $L^{(1)}(\mathcal{G}_j)$ in descending order to enable $\mathcal{G}_j$ to fill the bubbles of $\mathcal{G}_{j-1}/\mathcal{G}_{j+1}$ (Figure~\ref{fig:inter_stage_orch}(b)).
(2) Keeping micro-batches of the same bucket consecutive, as they are perfectly matched in terms of latency.
(3) Eagerly launching as many micro-batches as possible within memory limits (\S\ref{sec:data}) to ensure sufficient pending batches per stage~\cite{eager-1f1b}.
With these rules, \sysname{} enhances efficiency by $1.17\times$ while maintaining structured execution.
We have provided a detailed \textbf{optimality analysis} in Appendix~\S\ref{append-a} that demonstrates this structured pipeline always mitigates internal bubbles at the last stage, thereby achieving the near-optimal execution.

\subsubsection{Intra-Stage Orchestration}\label{sec:intra-stage}

\begin{figure}
\centering
\includegraphics[width=.98\linewidth]{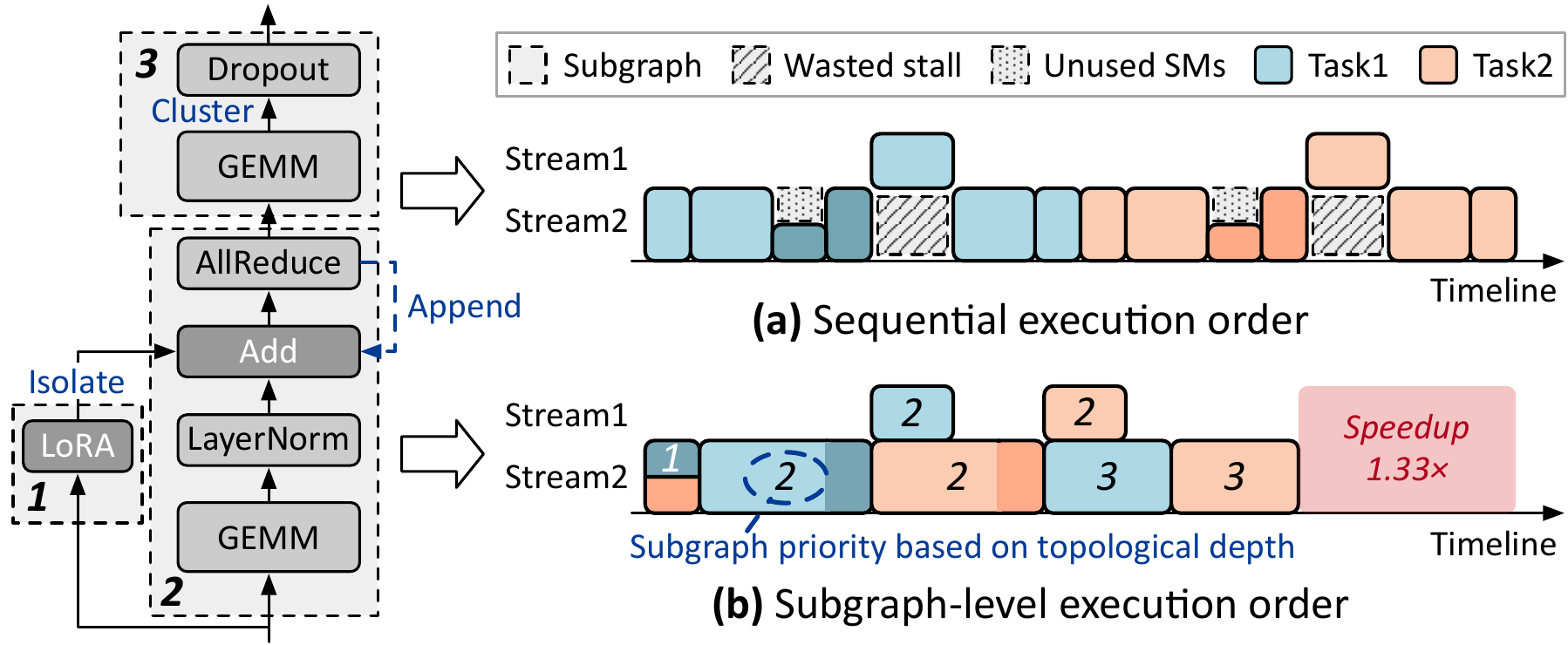}
\vspace{-1mm}
\caption{Example of intra-stage orchestration. Left is an example computational graph with darker blocks for adapters. 
}
\label{fig:intra_stage_orch}
\vspace{-4mm}
\end{figure}

To optimize intra-stage execution within a hTask bucket $\mathcal{G}_j$, we schedule computation and communication operators of hTasks $\{\mathcal{H}_i\} \in \mathcal{G}_j$ to mitigate device stalls, which is equivalent to minimizing stage latency as illustrated in Figure~\ref{fig:intra_stage_orch}.

\paragraph{Dependency-Aware Graph Construction.}
The computational graph of each hTask is a directed acyclic graph (DAG), naturally driving the intra-stage orchestration across hTasks as a multi-DAG scheduling problem~\cite{dag-sched}.
\sysname{} uses \textit{subgraph} as the minimal orchestrating unit, because model execution is sequential, while a longer sequence of computation operators benefits fully overlapping communication.

The subgraph construction is operated in a \textit{dependency-aware} manner.
\sysname{} first segments each DAG into subgraphs by clustering consecutive computation operators and appending each communication operator to its dependent operator (left part of Figure~\ref{fig:intra_stage_orch}).
Small-scale adapters are isolated as independent subgraphs.
Then, \sysname{} assigns a priority value to each subgraph according to its topological depth. 
This ensures interleaved execution while adhering to cross-graph dependencies created by adapter fusion (\S\ref{sec:adapter_fusion}).
Notably, the subgraph concept is compatible with vertical fusion-based kernels and CUDA graph techniques~\cite{nv-apex,cuda_graph}.


\paragraph{Subgraph Scheduling.}
Given multiple segmented DAGs, \sysname{} extends the Kahn algorithm~\cite{kahn-algo} to a multi-DAG variant in a latency-aware manner. 
Algorithm~\ref{alg:topology} outlines the scheduling process.
Given initial hTasks, it maintains a priority queue $\mathcal{P}$ to track zero in-degree subgraphs in each DAG (line 3-5).
In each iteration, the algorithm filters for the highest-priority subgraphs, selecting the one with the longest cumulative latency (from internal operators) to maximize overlap with in-flight communication operators. 
Then, the subgraph is removed from its DAG, while new zero in-degree ones are enqueued (line 6-13).
With \textit{launch\_schedule}, \sysname{} efficiently coordinates the runtime execution of multi-task subgraphs within the pipeline stage (\S\ref{sec:implement}).





\begin{algorithm}[t]
\begin{algorithmic}[1]
\small

\State \textbf{Input}: hybrid tasks $\mathcal{H} = \{\mathcal{H}_1, ...., \mathcal{H}_n\}$

\Function{\texttt{SubgraphSchedule}}{$\mathcal{H}$}
    \State $\mathcal{P} \leftarrow \texttt{PriorityQueue()}$ 
    \For{$\mathcal{H}_i \in \mathcal{H}$}   \Comment{\text{Initialize queue}}
    \State $\mathcal{P}.\texttt{enqueue}($\Call{\texttt{GetZIDSubgraphs}}{$\mathcal{H}_i.graph$}$)$
    \EndFor

    \State $launch\_schedule \leftarrow \emptyset; \ t \leftarrow 0$
    \While{$\mathcal{P} \neq \emptyset$}
    \State $subgraph \leftarrow \mathcal{P}.\texttt{dequeue(max\_lat)}$ \Comment{\text{Highest priority}}
    \State $DAG \leftarrow subgraph.parent\_graph$ \Comment{\text{Parent DAG}}
    \State $DAG.\texttt{remove}(subgraph)$
    \State $\mathcal{P}.\texttt{enqueue}($\Call{\texttt{GetZIDSubgraphs}}{$DAG$}$)$
    \State $launch\_schedule.\texttt{record}(<subgraph, \ t>)$
    \State $t \leftarrow t + subgraph.latency$ \Comment{\text{Update timer}}
    \EndWhile

    \State \Return{$\textit{launch\_schedule}$}

\EndFunction

\caption{Priority-Based Subgraph Scheduling}
\label{alg:topology}
\end{algorithmic}
\end{algorithm}

\subsubsection{Adapter Fusion and Communication Overlapping}\label{sec:adapter_fusion}

To improve GPU utilization of small-scale, PEFT-native operators, we adopt \textit{horizontal adapter fusion} with fine-grained management of GPU resources (\S\ref{sec:implement}), given that adapters cannot be directly batched across tasks.
The fusion strategy consists of three cases:
(1) For spatially batched tasks within the same hTask, their adapters are fused;
(2) For hTasks within the same bucket (i.e., interleaved in intra-stage parallelism), if all assigned a single task, their adapters are fused without impeding inter-task overlap.
As shown in Figure~\ref{fig:intra_stage_orch}, the \texttt{LoRA} operators of \texttt{Task1} and \texttt{Task2} are fusible because they are not in the same subgraph of \texttt{AllReduce} operator.
Conversely, \texttt{Add} operators cannot be fused, because the fusion would introduce a global synchronization in prior to the \texttt{AllReduce} operators of \texttt{Task1} and \texttt{Task2}.
(3) No fusion occurs across buckets (i.e., interleaved in inter-stage parallelism).

When overlapping computation and communication across different streams in intra-stage orchestration (Figure~\ref{fig:intra_stage_orch}), we observe that excessive CTA usage of communication kernels undermines compute kernel efficiency, while too few CTAs underutilize NVLink connection.
To resolve this tradeoff, we adopt NVLink SHARP~\cite{nvlink} to offload reductions into the NVSwitch, sustaining near-peak link bandwidth with a small CTA budget.
As a result, the network kernel fully overlaps with computation of other tasks using only 8 CTAs.

\subsection{Data Alignment}\label{sec:alignment}

\begin{figure}
\centering
\includegraphics[width=\linewidth]{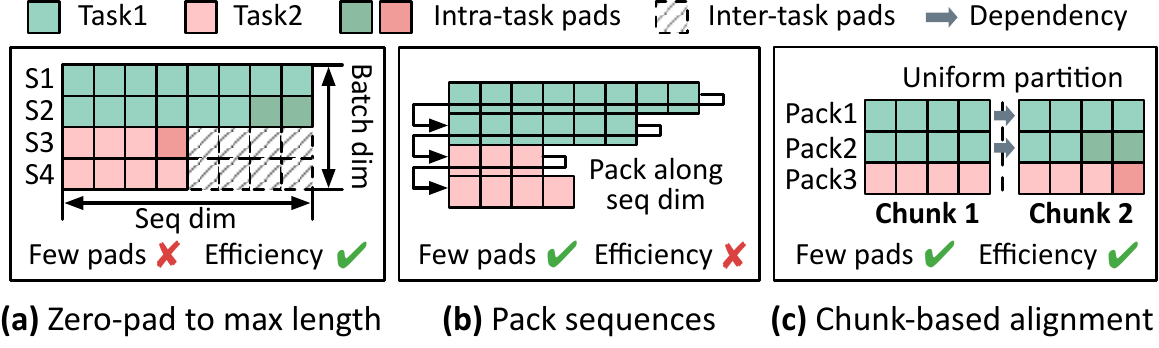}
\vspace{-6mm}
\caption{Example of existing data alignment strategies and our chunk-based alignment (\texttt{Si}: $i$-th sequence). \texttt{Dependency} represents KV cache reuse in causal attention computation.}
\label{fig:alignment}
\vspace{-2mm}
\end{figure}

To execute a hybrid task (hTask), the data batches of its internal PEFT tasks need to be aligned along sequence dimension.
One common strategy is to zero pad all sequences to a global maximum length (Figure~\ref{fig:alignment}(a)).
This incurs substantial inter-task ineffective tokens (i.e., without semantic information), thereby wasting compute and memory resources.
Another industrial-grade approach for pretraining is to pack sequences into longer ones.
However, recent works~\cite{bai2024longalign,dynapipe} observe that relying solely on packing degrades fine-tuning efficiency, as attention masks lead to wasted attention computation across sequences. 
Consequently, fine-tuning APIs often mandate sequence padding to maximum lengths, with intra-task zero-padded tokens billed to users~\cite{togetherai-peft}.
However, in \sysname{}, inter-task ineffective tokens arise from data alignment in co-scheduling and cannot be billed to users.
Therefore, \sysname{} focuses on \textit{mitigating inter-task ineffective tokens}.

\paragraph{Reinventing Packing with Chunk-Based Alignment.}
To mitigate inter-task padding, \sysname{} adopts a dual-step alignment strategy that maximizes efficiency without compromising model quality (Figure~\ref{fig:alignment}(c)).
First, it adaptively packs sequences within a single global batch for each task, respectively, to ensure no impact on model convergence (e.g., \texttt{Pack1}/\texttt{Pack2} for \texttt{Task1} and \texttt{Pack3} for \texttt{Task2}).
This step transforms per-task sequences into a set of longer, denser packed ones.
Second, \sysname{} uniformly partitions packed sequences into equal-sized chunks (e.g., size of 4 in Figure~\ref{fig:alignment}(c)).
For sequences longer than the chunk size (e.g., \texttt{Pack1}), \sysname{} scatters them across multiple consecutive chunks with the dependency of KV cache reuse~\cite{terapipe}.
This step not only mitigates redundant cross-sequence attention computation but also breaks overlong packed sequences into shorter ones for more fine-grained pipeline, which benefits throughput and peak activation memory reduction.

\begin{figure}
\centering
\includegraphics[width=.95\linewidth]{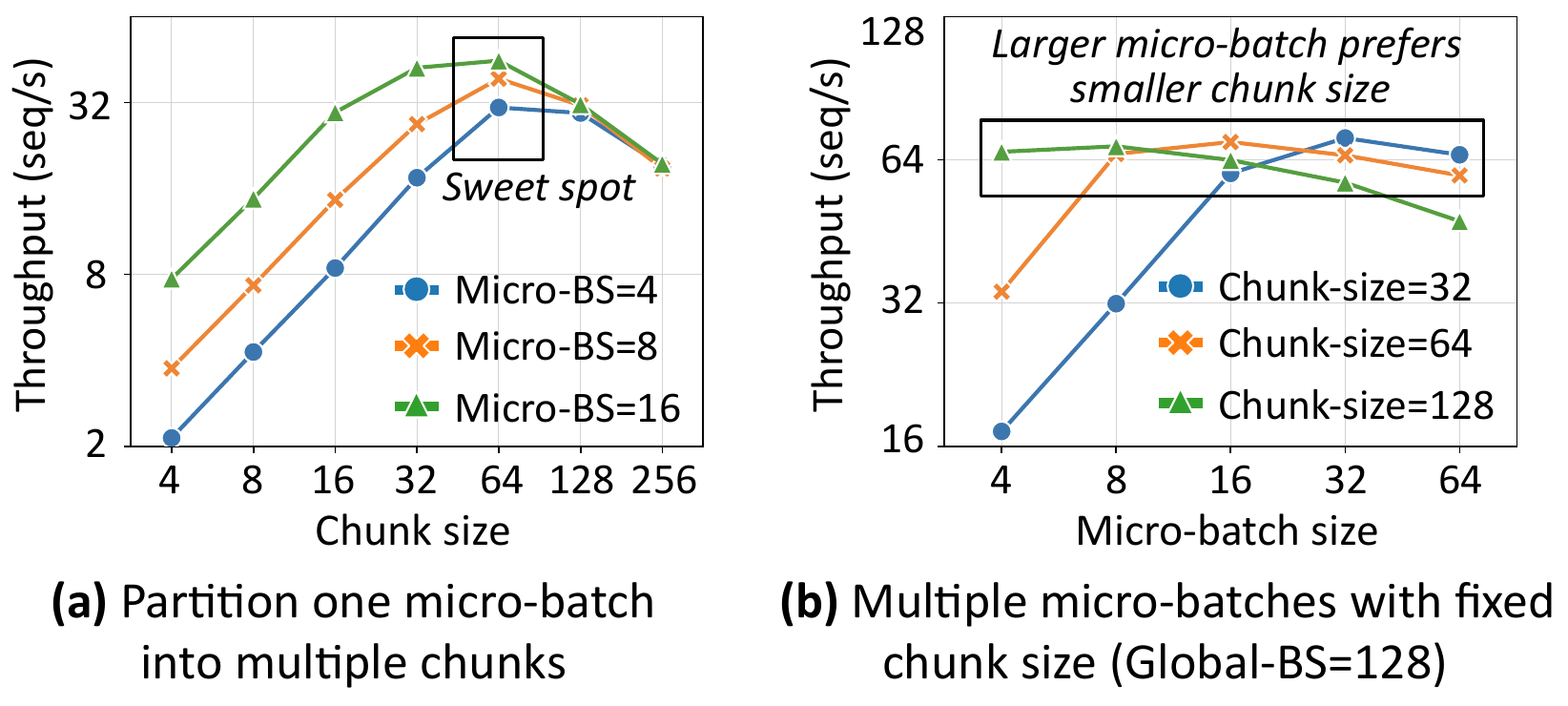}
\vspace{-2mm}
\caption{Quantifying chunk-based alignment (1 task, $16$-layer LLaMA7B, 4-GPU pipeline, sequence length 256).
}
\label{fig:chunk_tradeoff}
\vspace{-2mm}
\end{figure}

The determination of chunk size involves a tradeoff between compute efficiency and padding reduction, as depicted in Figure~\ref{fig:chunk_tradeoff}.
Smaller chunk size reduces padded tokens but risks underutilization and extra KV cache accesses; conversely, over-sized chunks hinder padding reduction and inflate stage latency. 
In practice, we set chunk size as the greatest power-of-2 divisor of all sequence lengths, with a minimum threshold (typically $64$) to avoid underutilization.




\section{Implementation}\label{sec:implement}

We build \sysname{} with 14K LOCs of Python based on Megatron-LM~\cite{megatron} and PyTorch~\cite{pytorch}, and 2K LOCs of C++ and CUDA for kernel implementation.
\sysname{} supports deployment with hybrid parallelism, including pipeline (e.g., GPipe~\cite{gpipe}, 1F1B~\cite{pipedream}, interleaved-1F1B~\cite{megatron-scale}), Megatron tensor parallelism~\cite{megatron}, and data parallelism strategies (e.g., PyTorch DDP~\cite{ddp}, FSDP~\cite{fsdp}).
\sysname{} implements dynamic \textit{Adapter} attachment to \textit{BaseOp} via PyTorch \textit{hook mechanism}~\cite{hook}, wrapping the logic of \textit{Dispatch} and \textit{Aggregate} sub-modules into the hooked \textit{BaseOp} function at runtime.

\paragraph{Grouped Kernels.}
We implement \sysname{}'s kernels based on NVIDIA CUTLASS and CuTe~\cite{cutlass} for grouped computation across task adapters.
In each kernel, we first assign thread blocks in proportion to the FLOPs and memory access of adapters for each task.
Then, we decompose each adapter operator into tiles, assigning those that access the same partial of operator weight to the same allocated thread block.
This implementation enhances load balancing across SMs while reducing   memory read/write overhead.


\paragraph{Subgraph Execution.}
We use TorchFX~\cite{torch-fx} to trace the PEFT model into multi-task IR graphs and conduct subgraph-level orchestration (\S\ref{sec:orchestration}).
Based on it, we adopt multiple streams and CUDA primitives (e.g., \texttt{Event.synchronize()}) to coordinate operator execution across independent tasks.


\paragraph{Offline Profiling and System Overhead.}
We conduct offline profiling across canonical operator configurations to build the cost model, given that hardware and backbones are pre-accessible.
This is backed by PyTorch \textit{dispatching mechanism} that ensures consistent kernel selection for identical input shapes, data types, and hardware~\cite{torch-dispatcher}.
We thus limit scheduling overhead to under 10 seconds by avoiding labor-intensive GPU operations, which is minimal compared to a fine-tuning task with a typical duration of 3-70 hours.

\section{Evaluation}\label{sec:eval}

\setlength{\tabcolsep}{3pt}
\begin{table}[t]
\small
\centering
\caption{Model configurations used in experiments. \texttt{\#GPUs} denotes the number of GPUs for each model unless specified.}
\vspace{-2mm}
\begin{tabular}{ccccc}
\toprule
Model & \#Layers  & Hidden Dim & \#Heads & \#GPUs\\
\midrule
GPT3-2.7B~\cite{gpt-3} & 32 & 2560 & 32 & 2\\
LLaMA2-7B/13B~\cite{llama2} & 32/40 & 4096/5120 & 32/40 & 4/8\\
OPT-30B~\cite{opt} & 48 & 7168 & 56 & 16\\
\bottomrule
\end{tabular}
\label{tab:models}
\vspace{-3mm}
\end{table}

\begin{figure*}
\centering
\includegraphics[width=\linewidth]{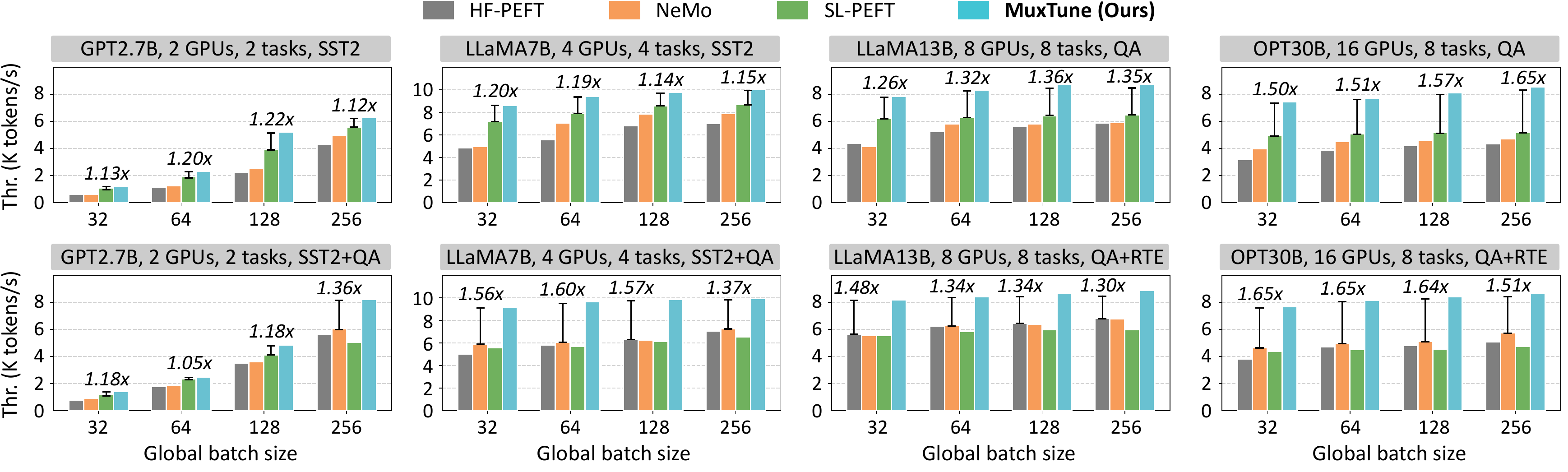}
\vspace{-6mm}
\caption{System throughput (in the number of processed tokens per second) across different global batch sizes, backbone models, and hardware configurations. Detailed workloads are presented above each subfigure (in grey blocks).
}
\label{fig:exp-e2e-thr}
\vspace{-3mm}
\end{figure*}

\subsection{Experimental Setup}\label{exp:setup}

\paragraph{Testbeds.}
We evaluate \sysname{} on three server setups: 
(1) Testbed-A: $1$ node with $4$ NVIDIA A40 GPUs ($48$GB), Intel Xeon Silver $4310$ CPU, and NVLink; 
(2) Testbed-B: $8$ nodes, each with $2$ NVIDIA A40 GPUs ($48$GB), Intel Xeon Gold 5318Y CPU, and Mellanox ConnectX-5~\cite{cx5} (100Gb/s infiniband). 
(3) Testbed-C: $1$ node with $8$ NVIDIA H100 GPUs ($80$GB), Intel Xeon Platinum 8457C CPU, and NVLink.

\paragraph{Models and Datasets.}


We conduct experiments with four representative LLMs in Table~\ref{tab:models}. 
We have implemented three PEFT types: LoRA~\cite{lora} (mainly used), Adapter Tuning~\cite{adapter-tuning} and Diff Pruning~\cite{diff-pruning}.
We use three datasets with varied sequence length, including SST2~\cite{sst2}, OpenBookQA~\cite{openbookqa}, and RTE~\cite{rte}.
Since sequences of each task are padded to its maximum length, we analyze dataset distribution and pad (or truncate) sequences of SST2 to $64$, QA to $128$, and RTE to $256$.
In evaluation, we use two dataset combinations (\S\ref{sec:peft_basic}): 
(1) \texttt{Uniform} shares the same dataset across tasks colocated on the same backbone; 
(2) \texttt{Non-uniform} adopts different datasets for these tasks.
The testbed scales and workloads are sufficient to evaluate system performance, because PEFT consumes much less memory than pretraining while featuring limited input sizes (thus no large data parallelism is needed).

\paragraph{Baselines.}

We compare \sysname{} with three baselines:
\begin{enumerate}[label=(\arabic*), itemsep=0pt, parsep=0pt, labelsep=2pt, leftmargin=*, topsep=2pt,partopsep=0pt]
  \item \textit{HuggingFace PEFT}~\cite{hf_peft} (HF-PEFT): a user-friendly library for adapting various LLMs to downstream tasks, with support for memory optimizations like quantization.

  \item \textit{NeMo Megatron}~\cite{nemo} (NeMo): an AI framework built on Megatron-LM~\cite{megatron} that supports efficient kernels and scalable parallelism strategies to adapt LLMs with PEFT.

  \item \textit{SLoRA-PEFT}~\cite{slora} (SL-PEFT): It adopts SLoRA's techniques like backbone sharing and batching-only in PEFT, supporting fine-tuning multiple tasks concurrently.
\end{enumerate}

\paragraph{Parallelism Selection.}
We grid-search the optimal parallelism for \sysname{} and baselines across supported strategies (\S\ref{sec:implement}).
For 2-GPU and 4-GPU cases, the search locates intra- and inter-stage parallelism, respectively.
For more GPUs with heavier workloads, hybrid parallelism is selected (intra-stage parallelism within each node and pipeline across nodes).

\subsection{End-to-End Performance}\label{exp:e2e}


\begin{figure}
\centering
\includegraphics[width=.98\linewidth]{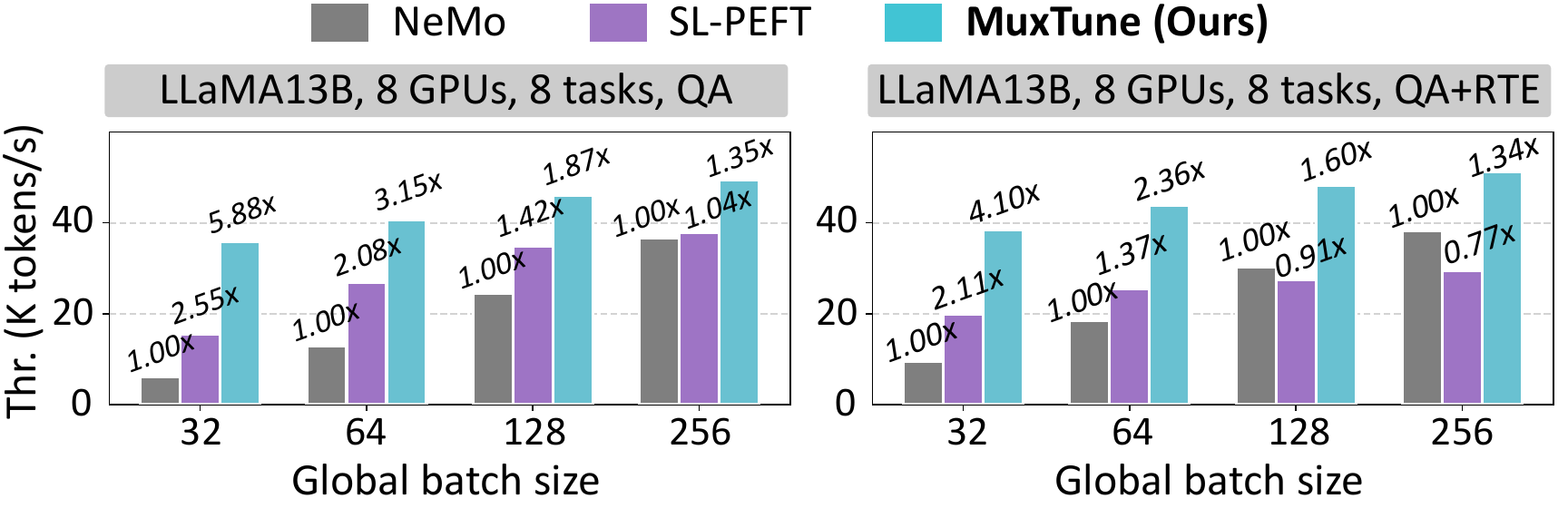}
\vspace{-2mm}
\caption{Throughput on H100 GPUs across global batch sizes. Configurations are aligned with the above experiment.
}
\label{fig:exp-h100}
\vspace{-2mm}
\end{figure}

\paragraph{System Throughput.}
Figure~\ref{fig:exp-e2e-thr} demonstrates the end-to-end throughput of \sysname{} and baseline systems across various workloads.
In the \texttt{Uniform} case, \sysname{} improves throughput by up to $2.33\times$, $1.87\times$, and $1.64\times$ over HF-PEFT, NeMo, and SL-PEFT, respectively.
With lightweight workloads, \sysname{} surpasses baselines by adaptively spatially fusing multiple tasks to improve utilization.
The performance gains grow with more GPUs, benefited from the ability of \sysname{} to overlap multi-task operators for stall reduction.

In the \texttt{Non-uniform} case, \sysname{} achieves throughput improvements of $2.23\times$, $1.83\times$ and $1.85\times$ over the three baselines, respectively.
For HF-PEFT and NeMo, the improvement remains consistent with \texttt{Uniform} case, as they execute tasks separately without extra zero-padded tokens.
For SL-PEFT, \sysname{} achieves higher improvement than in the \texttt{Uniform} case, because SL-PEFT incurs substantial zero-padded tokens that waste compute and memory resources.

\paragraph{Performance on Advanced Hardware.}
We further evaluate \sysname{} against two baselines on H100 GPUs.
As illustrated in Figure~\ref{fig:exp-h100}, \sysname{} improves throughput by $5.29\times$ and $2.31\times$ over NeMo and SL-PEFT in the \texttt{Uniform} case (left), and $3.69\times$ and $1.94\times$ in the \texttt{Non-uniform} case (right).
Beyond the above reasons, the more significant performance gains on H100 (compared to A40) stem from its superior compute power, which amplifies the underutilization inherent in single-task PEFT frameworks while unlocking more multi-task optimization potential for \sysname{}.

\subsection{Ablation Studies}\label{exp:ablation}

\begin{figure}
\centering
\includegraphics[width=0.98\linewidth]{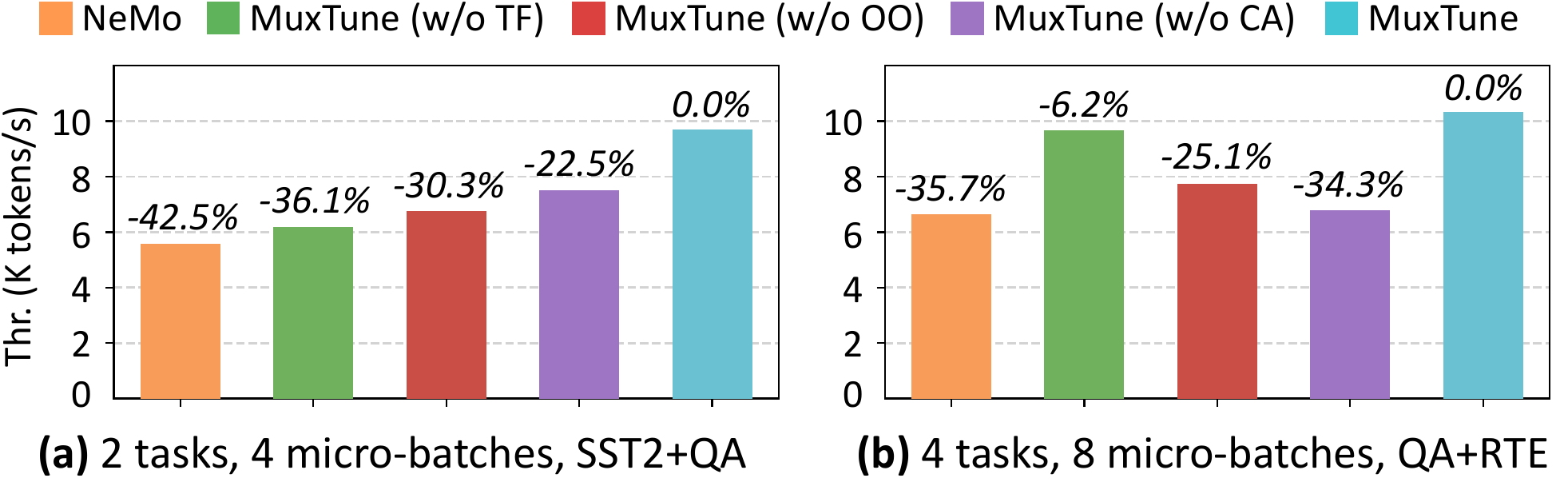}
\vspace{-2mm}
\caption{Performance breakdown. \texttt{TF}, \texttt{OO} and \texttt{CA} represent task fusion, operator orchestration, and data alignment.}
\label{fig:exp-ablation}
\end{figure}

\paragraph{Performance Breakdown.}
Figure~\ref{fig:exp-ablation} demonstrates the impact of each component in \sysname{} using LLaMA7B, $4$-GPU pipeline, and global batch size of 128.
With lightweight workloads (Figure~\ref{fig:exp-ablation}(a)), disabling three components undermines throughput by $36.1\%$, $30.3\%$, and $22.5\%$, respectively.
This arises from improved GPU utilization (from computation fusion), reduced device stalls (from communication overlapping), and mitigated ineffective tokens across tasks (from chunk-based alignment).
With heavier workloads (Figure~\ref{fig:exp-ablation}(b)), data alignment increasingly dominates the overall throughput, causing a $34.3\%$ decrease.
Conversely, disabling task fusion only undermines throughput by $6.25\%$ as GPU has been saturated. 
Since more micro-batches lead to fewer device stalls, disabling operator orchestration results in a $25.1\%$ drop, slightly less than with lightweight workloads.



\setlength{\tabcolsep}{3pt}
\begin{table}[t]
\small
\centering
\caption{Task workloads (WL) with random generated configurations (dataset, batch size) used in experiments.
}
\vspace{-2mm}
\begin{tabular}{c|cccccccc}
\toprule
{Task Order} & {\#1}  & {\#2} & {\#3} & {\#4} & {\#5}  & {\#6} & {\#7} & {\#8} \\
\midrule
\text{WL-A} & SST2 & QA & QA & SST2 & SST2 & SST2 & QA & QA \\
\text{WL-B} & RTE & SST2 & RTE & SST2 & SST2 & RTE & RTE & RTE \\
\addlinespace[1pt]
\hline
\addlinespace[1pt]
{Batch Size} & 4 & 2 & 4 & 4 & 8 & 2 & 4 & 4 \\
\bottomrule
\end{tabular}

\label{tab:workloads}
\vspace{-1mm}
\end{table}

\begin{figure}[t]
\centering
\includegraphics[width=.92\linewidth]{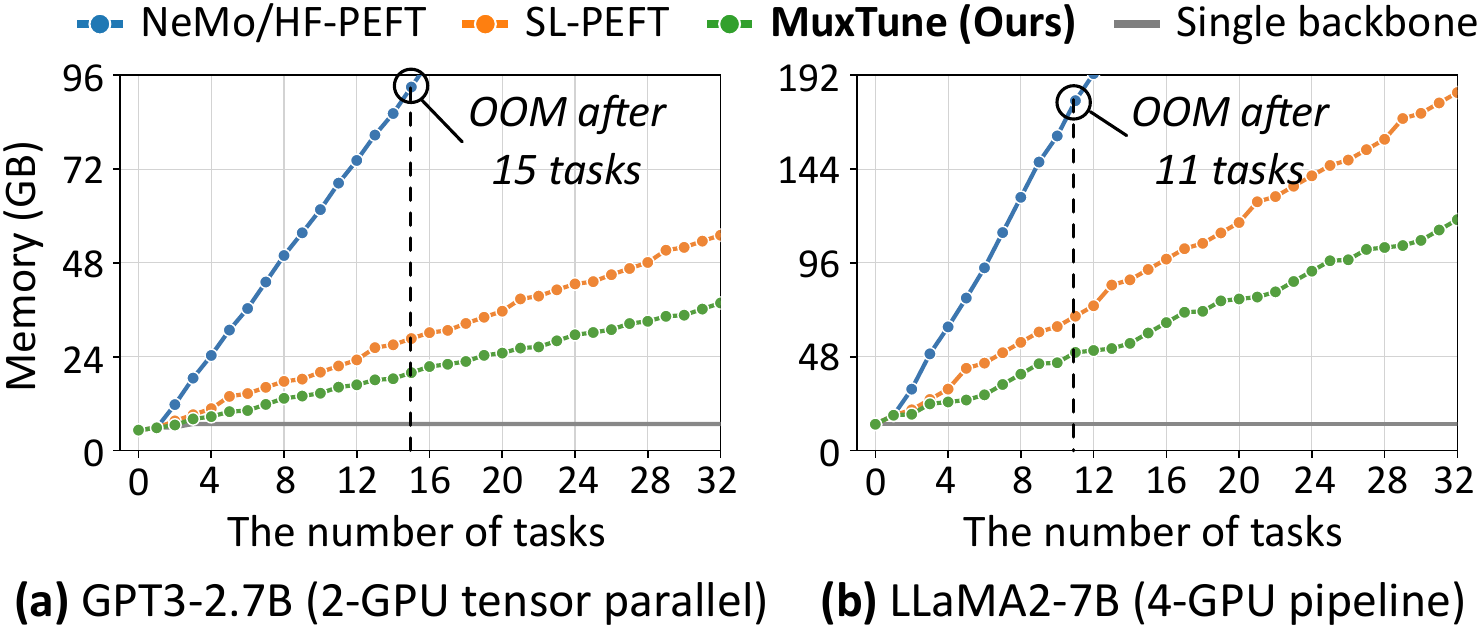}
\vspace{-2mm}
\caption{Memory footprint with various number of tasks.
}
\label{fig:exp-mem}
\vspace{-1mm}
\end{figure}

\paragraph{Memory Efficiency Analysis.}
We exclusively evaluate the memory efficiency of \sysname{} against other baselines using two PEFT task workloads as listed in Table~\ref{tab:workloads}. 
As shown in Figure~\ref{fig:exp-mem}, we submit 32 PEFT tasks in a progressive manner, each with 1 micro-batch per iteration, by repeating the workloads four times.
We employ the GPT2.7B backbone for WL-A and LLaMA7B for WL-B, which incur memory consumption of $5.2$GB and $13.4$GB, respectively.

In Figure~\ref{fig:exp-mem}(a), with tensor parallelism on 2 A40 GPUs (48GB each), \sysname{} achieves up to $4.67\times$ and $1.44\times$ memory reduction over NeMo/HF-PEFT (OOM after 15 tasks) and SL-PEFT, respectively.
Without memory constraints (i.e., scaling up to 32 tasks), \sysname{} further reduces memory footprint by $5.29\times$ and $1.46\times$ compared to these baselines.
The reasons are two-fold: 
(1) \sysname{} flexibly shares the memory-intensive backbone across tasks, which consumes 13.4GB in contrast to 4.3GB of activation and 0.4GB of others (Figure~\ref{fig:exp-mem}(b)), while NeMo and HF-PEFT replicate one backbone per task; 
(2) \sysname{} mitigates inter-task ineffective tokens and benefits fine-grained pipeline (\S\ref{sec:alignment}), while SL-PEFT excessively batches tasks with substantial padding and higher peak activation memory.
For larger backbone and more GPUs (Figure~\ref{fig:exp-mem}(b)), \sysname{} reduces memory footprint by $3.57\times$ and $1.37\times$ over NeMo/HF-PEFT (OOM after 11 tasks) and SL-PEFT, demonstrating better scalability of \sysname{} against baseline systems.

\begin{figure}[t]
\centering
\includegraphics[width=\linewidth]{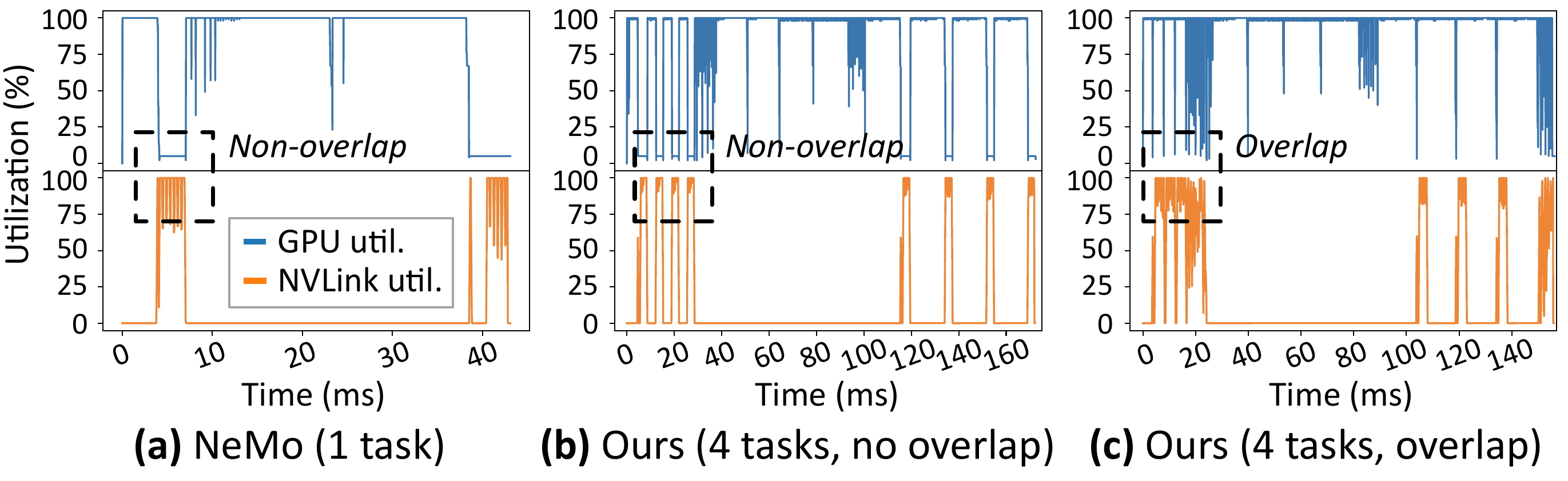}
\vspace{-6mm}
\caption{GPU and NVLink utilization of 1 layer with $4$-GPU tensor parallelism. 
Tasks are interleaved in (b) and (c).}
\label{fig:exp-gpu-util}
\vspace{-1mm}
\end{figure}

\paragraph{Efficiency of Operator Orchestration.}\label{exp:op_orchest}
We visualize the GPU compute and network bandwidth utilization of \sysname{} with comparison to the NeMo baseline.
Figure~\ref{fig:exp-gpu-util} uses Nsight toolkit~\cite{nsys} to profile the GPU and NVLink utilization under tensor parallelism.
As shown in Figure~\ref{fig:exp-gpu-util}(a), NeMo executes a single task with sequentially launched operators.
Since computation is blocked by communication, the average GPU utilization remains at $82.5\%$, with a layer latency of $43.2$ms.
Figure~\ref{fig:exp-gpu-util}(b) shows the results of $4$ tasks with interleaved execution, but without overlap.
The latency linearly increases to $172.5$ms, while GPU utilization remains nearly constant at $84.7\%$.
In Figure~\ref{fig:exp-gpu-util}(c), \sysname{} fully overlaps computation with communication.
Without being blocked, the GPU utilization reaches $97.8\%$, achieving a $1.19\times$ improvement over the baseline, and the $4$-task latency of the decoder layer is reduced to $156.2$ms. 

\begin{figure}[t]
\centering
\includegraphics[width=.98\linewidth]{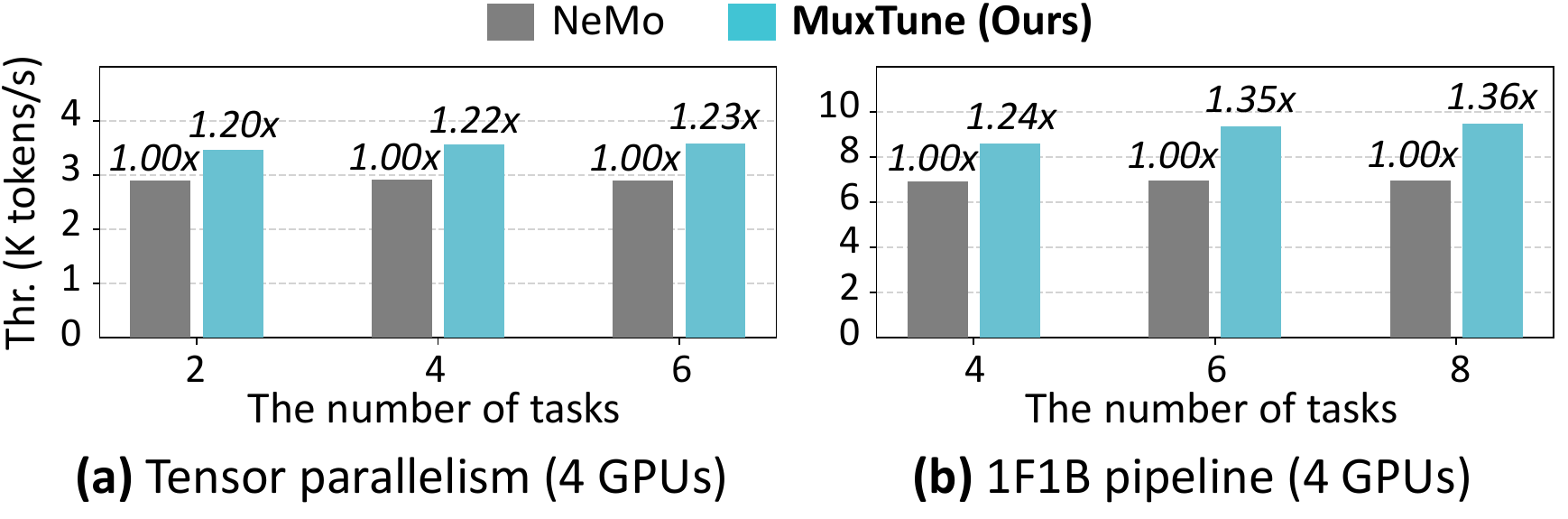}
\vspace{-2mm}
\caption{Throughput of operator orchestration with varying number of tasks (LLaMA7B, sequence length $128$, $64$, $32$). 
(a) 1 micro-batch of size 8, (b) 8 micro-batches of size 8.
}
\label{fig:exp-stall-reduce}
\vspace{-1mm}
\end{figure}

We also evaluate the end-to-end performance across different parallelism strategies and workloads, as shown in Figure~\ref{fig:exp-stall-reduce} with only backbone sharing and operator orchestration enabled.
Figure~\ref{fig:exp-stall-reduce}(a) illustrates the benefits of inter-task overlapping in tensor parallelism.
With an increasing number of tasks, \sysname{} delivers $1.20\times$, $1.22\times$, and $1.23\times$ higher throughput than NeMo, owing to the overlap between computation and communication.
Figure~\ref{fig:exp-stall-reduce}(b) demonstrates the effect of pipeline orchestration.
Compared to NeMo, \sysname{} improves throughput by $1.24\times$, $1.35\times$, and $1.36\times$, as the interleaved stage computations across tasks effectively mitigate pipeline bubbles.
Notably, \sysname{} is capable of achieving higher improvements (e.g., $1.59\times$ with 4 micro-batches) with fewer micro-batches (more pipeline bubbles).

\paragraph{Effectiveness of Data Alignment.}\label{exp:alignment}
We assess chunk-based data alignment using the metric \textit{effective throughput}~\cite{goodput}, which measures the throughput of original tokens excluding inter-task zero-padded ones, reflecting the economic gains of service providers (\S\ref{sec:alignment}).
We use workloads in Table~\ref{tab:workloads} to evaluate cases with and without intra-chunk zero-padding.

\begin{figure}[t]
\centering
\includegraphics[width=.98\linewidth]{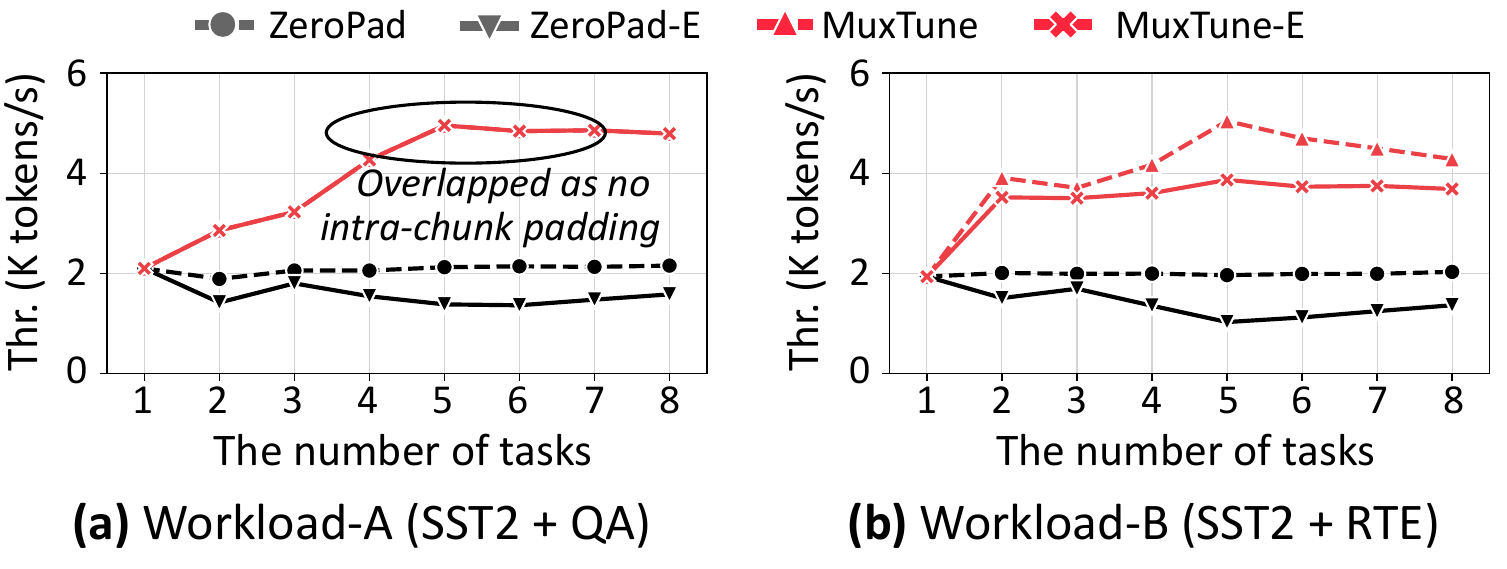}
\vspace{-2mm}
\caption{Throughput of 1 hybrid task with various number of tasks and datasets (LLaMA7B, $4$-GPU pipeline). 
\texttt{ZeroPad} represents zero-padding all sequences as in SL-PEFT. \texttt{-E} denotes effective throughput, and overall if not marked.
}
\label{fig:exp-align-effect}
\vspace{-1mm}
\end{figure}

Figure~\ref{fig:exp-align-effect} demonstrates the throughput of progressively adding tasks into a single hybrid task with one micro-batch.
In Figure~\ref{fig:exp-align-effect}(a) with a chunk size of 64 (matching SST2), \sysname{} avoids intra-chunk padding and achieves up to $2.33\times$ higher throughput and $3.59\times$ higher effective throughput than SL-PEFT.
This is because \sysname{} effectively mitigates inter-task padding without underutilizing GPUs.
In Figure~\ref{fig:exp-align-effect}(b) with more inclined sequence lengths, \sysname{} introduces intra-chunk zero-padding for SST2 tasks with a chunk size of $128$.
In this case, \sysname{} still achieves up to $3.77\times$ improvement on overall throughput, and $2.57\times$ on effective throughput compared to SL-PEFT.

\subsection{Scalability and Scheduling Study}\label{exp:scalability}

\begin{figure}[t]
\centering
\includegraphics[width=.98\linewidth]{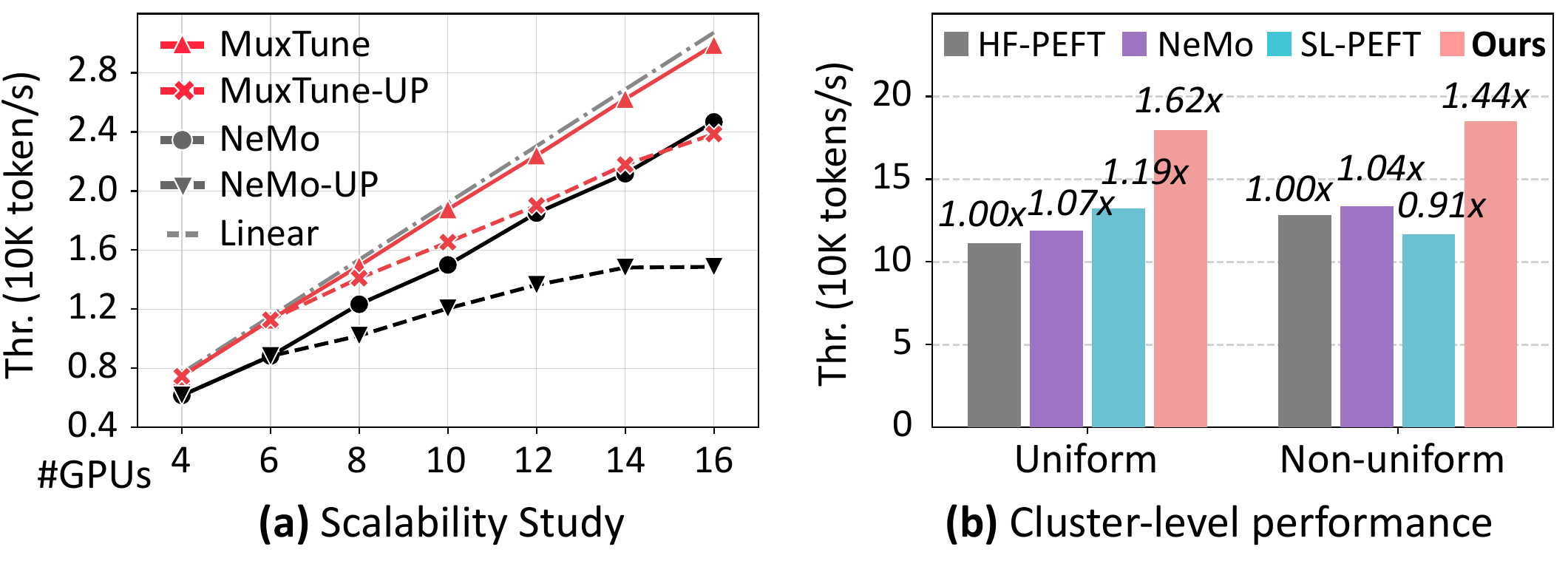}
\vspace{-2mm}
\caption{System scalability of \sysname{} and cluster-level performance under production-grade workloads.}
\label{fig:exp-scale-sched}
\vspace{-1mm}
\end{figure}

\paragraph{Scalability Study.}
Figure~\ref{fig:exp-scale-sched}(a) demonstrates the system scalability with two scaling strategies (LLaMA7B, global batch size 128, micro-batch size 8, $n$ tasks for $n$-GPU): 
(1) ``Up-only'' (\texttt{-UP}): Scales up (i.e., increases allocated GPUs) for active instances as workloads increase;
(2) ``Up-then-out'': Scales up first, then scales out (i.e., replicates new instances) if workloads continue increasing.
For the ``up-only'' strategy, despite sub-linear scaling (\S\ref{sec:data}), \sysname{} delivers $1.61\times$ higher throughput than NeMo by efficiently improving utilization and mitigating device stalls.
For the ``up-then-out'' strategy, both systems achieve near-linear scaling, while \sysname{} still delivering up to $1.28\times$ higher throughput.

\paragraph{Cluster-Level Performance.}
We further integrate \sysname{} into cluster scheduling and evaluate it under production-grade workloads.
In absence of public PEFT traces, we adapt a one-week Philly trace~\cite{jeon2019analysis}.
The average task duration and standard deviation are 372.6 min and 612.9 min, respectively, while the average arrival rate is 2.59 tasks/min.
We replay PEFT workloads in a simulated 128‑GPU cluster with a first‑come, first‑served scheduler, using both \texttt{Uniform} and \texttt{Non-uniform} combinations, LLaMA7B backbone, and randomly generated configurations for each task.
As shown in Figure~\ref{fig:exp-scale-sched}(b), in the \texttt{Uniform} case, \sysname{} enhances cluster throughput by $1.61\times$, $1.51\times$, and $1.36\times$ over HF-PEFT, NeMo, and SL-PEFT baselines, respectively.
In the \texttt{Non-uniform} case, \sysname{} further delivers $1.58\times$ higher cluster throughput against SL-PEFT, which highlights the effectiveness of chunk-based data alignment when fusing multiple tasks under workloads with variable-length sequences.


\section{Discussion and Future Work}

\paragraph{Generality to Cluster Scheduling Policies.}
Beyond the first-come, first-served scheduler used in evaluation (\S\ref{exp:scalability}), the system design of \sysname{} is extensible as the fine-tuning backend for other cluster scheduling policies, such as budget-based~\cite{kube}, task priority-based~\cite{gavel}, and SLO-aware scheduling~\cite{elasticflow}.
For instance, with task priorities, the scheduler can colocate low-priority tasks to boost instance-level throughput while allocating dedicated resources for high-priority ones to guarantee task-level latency.
Moreover, tasks with the same backbone type could be colocated for backbone multiplexing while those with different types should be scheduled to different instances.
We leave the in-depth exploration of ``multiplexing-aware'' scheduling as future work.



\paragraph{Extensibility to Performance Metric Optimizations.}
The optimizations of \sysname{} focus on maximizing instance-level throughput by enhancing resource efficiency and reducing per-task memory consumption. 
Beyond that, \sysname{} can achieve higher energy efficiency by mitigating wasted device stalls and lowering overall elapsed times of all colocated tasks~\cite{perseus}.
Users can extend our techniques to optimize other performance metrics, such as integrating admission control (i.e., limiting the number of tasks dispatched to fine-tuning instances) to guarantee all colocated tasks can be completed within their user-specified SLOs.
Moreover, users can adaptively scale the hardware frequencies while adhering to SLO requirements to further enhance energy efficiency~\cite{perseus,miu-serve}.


\section{Related Work}\label{sec:related}

\paragraph{Multi-Adapter LLM Systems.}

Recent years have witnessed significant progress in multi-adapter LLM systems~\cite{punica,slora,dlora-osdi,ymir,pets}.
PetS~\cite{pets} proposes a unified framework for concurrent PEFT task serving.
Punica~\cite{punica}, SLoRA~\cite{slora}, and dLoRA~\cite{dlora-osdi} focus on efficient scheduling and kernel implementation for multi-LoRA serving systems.
Unlike these works, \sysname{} targets multi-task execution optimization in PEFT scenarios, leveraging hierarchical co-scheduling to maximize resource utilization and minimize device stalls.

\paragraph{Techniques for Improving GPU Utilization.}

Several studies have explored GPU utilization optimizations through resource allocation and management across multiple ML applications, including temporal~\cite{cgpu, vcuda} and spatial sharing~\cite{mig, mps, cuda_stream} to multiplex the computational units.
Other works~\cite{horizontal_fusion,transformer_engine} horizontally or vertically fuse small kernels into larger ones to fully utilize hardware.
However, they fail to be directly applied in PEFT, due to the memory-intensive backbone (which limits scalability), the lack of operator-level execution control (which incurs inter-task interference), and the interdependencies between adapters for the same task.

\paragraph{Frameworks for LLM Parallelization and Reducing Device Stalls.}

The field of LLM parallelization has been intensively studied in recent years, including pipeline ~\cite{gpipe,pipedream}, tensor~\cite{megatron}, and data parallelism~\cite{zero_offload,zero}, as well as automatic parallelism optimizations~\cite{alpa,zero}.
To resolve the issue of device stalls introduced by these parallelism strategies, DeepSeek-V3~\cite{deepseekv3} and ZeroBubble~\cite{zerobubble} split the backward pass to reduce pipeline bubbles.
Overlapping-based methods~\cite{transformer_engine,wang2022overlap} decompose computations and overlap them with communication.
TeraPipe~\cite{terapipe} pipelines token-level computations to optimize long-sequence training.
These techniques are unsuitable for PEFT, owing to the absence of weight gradient computation and the potential of GPU underutilization.

\section{Conclusion}

\sysname{} is a resource-efficient system that optimizes concurrent multi-task PEFT execution in multi-tenant datacenters.
The core of \sysname{} is to multiplex the backbone across independent tasks in a spatial-temporal manner.
\sysname{} modularizes PEFT tasks for flexible backbone sharing, and devises hierarchical multi-task co-scheduling scheme with task, operator, and data-level optimizations to improve GPU utilization and reduce device stalls.
Experimental results show that \sysname{} achieves up to $2.33\times$ higher throughput and $5.29\times$ memory reduction compared to three baselines.

\section*{Acknowledgments}

We would like to thank the anonymous reviewers and our shepherd, Jiarong Xing, for their valuable feedback. 
This work is supported by the National Natural Science Foundation of China (62232011) and the Natural Science Foundation of Shanghai Municipality (24ZR1430500).
Quan Chen is the corresponding author of this paper (\href{mailto:chen-quan@cs.sjtu.edu.cn}{chen-quan@cs.sjtu.edu.cn}).

\bibliographystyle{plain}
\bibliography{references}

\clearpage
\appendix

\section{Optimality Analysis for Structured Pipeline Template}~\label{append-a}

Theoratically proving the optimality of a pipeline schedule is non-trivial, not to mention that micro-batches are heterogeneous, i.e., with different micro-batch sizes and sequence lengths.
Exhaustively enumerating all possible candidate pipeline schedules can be formulated as an integer linear programming (ILP) problem with $S\sum_i n_i$ constraints, which is proven to be NP-hard~\cite{karp2009reducibility} with the time complexity of $e^{O(S\sum_i n_i)}$, where $n_i$ is the number of micro batches of task $i$.
However, in our PEFT scenarios, micro-batches of each hTask bucket retain a consistent shape, while its forward and backward passes share the same latency for each pipeline stage (\S\ref{sec:inter-stage}).
These characteristics offer us opportunities to reduce the complexity of the problem.
Below, we theoretically discuss the optimality of our proposed pipeline template.

Our discussion focuses on how to achieve multi-task pipeline optimality (i.e., minimized end-to-end latency) in the paradigm of 1F1B~\cite{pipedream}, as it is one of the most widely used and efficient pipeline.
The execution of 1F1B pipeline can be divided into three phases, including warm-up, steady, and drain phase.
In multi-task 1F1B pipeline, we follow~\cite{pipedream,dynapipe} to first introduce a basic lemma: 
\begin{lemma}
    The end-to-end pipeline latency is calculated by adding the latencies of three phases ($T_{warm}$, $T_{steady}$, $T_{drain}$), while $T_{warm}$/$T_{drain}$ can be calculated as: $(S - 1)t_{1}$/$(S - 1)t_{P}$.
\label{lemma-1}
\end{lemma}

\begin{figure}
\centering
\includegraphics[width=\linewidth]{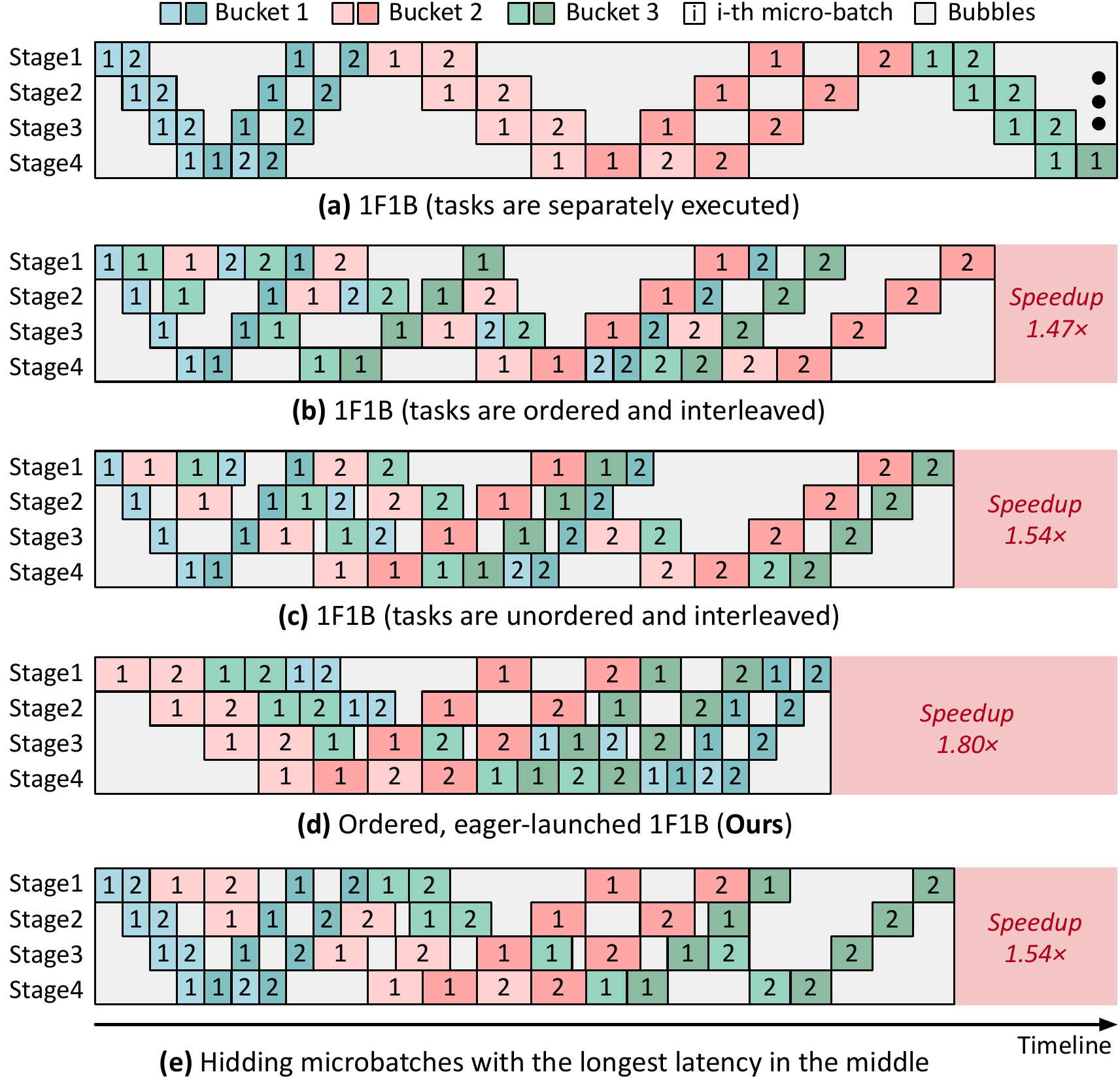}
\vspace{-6mm}
\caption{Various multi-task 1F1B pipeline schedules.
}
\label{fig:appendix-a}
\vspace{-4mm}
\end{figure}

In the above lemma, $C$ is the number of micro-batches, $P$ is the number of hybrid task (hTask) buckets, and $S$ is the number of pipeline stages. $t_i$ denotes the stage latency of the micro-batches for the $i$-th bucket. 
$t_{1}$ and $t_P$ represent the forward (backward) stage latency of the first and last sorted hTask bucket, respectively.
Then, as observed in Figure~\ref{fig:appendix-a}(b) and (c), we give the next lemma as follows:
\begin{lemma}
    In steady phase, all micro-batches are bound to go through the last stage for one forward and one backward pass, i.e., $T_{steady} \geq 2C\sum_{i=1}^P t_i$. 
\label{lemma-2}
\end{lemma}

After that, we can measure the latency ratio between $T_{steady}$ and $T_{warmup} + T_{drain}$ as:
\begin{equation}
\begin{aligned}
\frac{T_{steady}}{T_{warmup} + T_{drain}} \geq \frac{2C \sum_{i=1}^Pt_i }{(S-1)(t_1 + t_P)}.
\label{eq:appendix-a}
\end{aligned}
\end{equation}

Since we construct hTask buckets in a workload-balanced manner (i.e., minimizing inter-bucket variance), here we simplify the equation by assuming $t_i = t_{i-1}=t, \ \forall i \in [2, P]$:
\begin{equation}
\begin{aligned}
\frac{T_{steady}}{T_{warmup} + T_{drain}} \geq \frac{2CPt}{2(S - 1)t} = \frac{CP}{S-1}.
\label{eq:appendix-2}
\end{aligned}
\end{equation}

In common practice, the number of micro-batches is set to much larger than $S$ to reduce the pipeline bubble ratio (e.g., $4\times$ in GPipe~\cite{gpipe} and Alpa~\cite{alpa}).
Besides, in our scenario of multi-task backbone sharing, the number of hTask buckets is also typically much larger than $S$.
For example, in the ``Memory Efficiency Analysis'' of \S\ref{exp:ablation}, 4 A40 GPUs can accommodate more than $32$ PEFT tasks co-located on a single LLaMA7B backbone with 4-GPU pipeline.
That is, the number of hTask buckets $P$ can be set to up to $32$ when $S=4$.
As a result, we given the theorem as follows:
\begin{theorem}
    In multi-task PEFT scenario, the latency of steady phase ($T_{steady}$) typically dominate the end-to-end pipeline latency.
\label{theorem-1}
\end{theorem}

Therefore, to prove the near optimality of our proposed pipeline template, \textbf{we only need to prove that $T_{steady}$ is minimized to $2C\sum_{i=1}^P t_i$, i.e., no pipeline internal bubbles exist in the last stage, as shown in Figure~\ref{fig:appendix-a}(d)}.

To prove this statement, we notice that our template generation requires sorting hTask buckets $\mathcal{G} = \{\mathcal{G}_1,\mathcal{G}_2,...,\mathcal{G}_P\}$ by their latencies in descending order (micro-batches of the same bucket are consecutive), while advocating eagerly launching as many as micro-batches as possible.
In this context, we give the third lemma as follows:
\begin{lemma}
    Before the backward pass of $\mathcal{G}_j$'s last micro-batch in the last stage completes, the forward pass of $\mathcal{G}_{j+1}$'s first micro-batch in the second-to-last stage always have been ready.
\label{lemma-3}
\end{lemma}

As shown in Figure~\ref{fig:appendix-a}(d), this lemma is justified because: 
(1) forward and backward passes share identical latency, 
(2) the stage latency of $\mathcal{G}_{j+1}$ is always shorter than that of $\mathcal{G}_j$, and 
(3) 1F1B pipeline always prioritizes the ready backward passes.
With this lemma, we give the following theorem:
\begin{theorem}
    In our proposed pipeline template, once the first micro-batch begins its forward pass in the last stage, the last stage would ``keep busy'' until the backward pass of the last micro-batch completes.
\label{theorem-2}
\end{theorem}

Therefore, \textbf{we have completed the near optimality proof for our proposed pipeline template.}
It should be noted that the optimality proof is built on the premise that \textit{device memory is always sufficient for our micro-batch eager launching scheme}, since the memory footprint is greatly reduce in multi-task PEFT scenarios (\S\ref{exp:ablation}). 
In practice, we implement our pipeline template generation with memory limitations, i.e., delaying micro-batch launching if our memory cost model reports that at least one stage would incur OOM issues.

In Figure~\ref{fig:appendix-a}(e), we further illustrate the effect of hidding the longest micro-batches in the middle (rather than sorting in descending order).
As observed, despite the reduction of $T_{warmup}$ and $T_{drain}$, such modification disrupts Theorem~\ref{theorem-2} and thus leads to worse end-to-end pipeline latency.

\end{document}